\documentclass[12pt]{iopart}

\usepackage{color}
\usepackage[]{graphicx}  
\usepackage{cite}
\usepackage{amssymb}
\usepackage{amsfonts}
\usepackage{bm}

\newcommand{\bfE}{\mathbf{E}} %{\bi{E}}
\newcommand{\bfD}{\mathbf{D}} %{\bi{D}}
\newcommand{\bfd}{\mathbf{D}} %{\bi{d}}
\newcommand{\bfr}{\mathbf{r}} %{\bi{r}}
\newcommand{\bfn}{\mathbf{n}} %{\bi{n}}
\newcommand{\bfrho}{\boldsymbol{\rho}}
\newcommand{\cF} {\mathcal{F}}
\newcommand{\ext}{\mathrm{ext}}
\newcommand{\tot}{\mathrm{tot}}
\newcommand{\rd}{\mathrm{d}}
\newcommand{\re}{\mathrm{e}}
\newcommand{\ri}{\mathrm{i}}

\begin{document}
\title{Dynamical screening of an endohedral atom}

\author{S Lo$^1$, A~V Korol$^{1,2}$ and A~V Solov'yov$^1$}
\address{$^1$ Frankfurt Institute for Advanced Studies, %Johann Wolfgang
Goethe-Universit\"at, Ruth-Moufang-Str.~1, 60438 Frankfurt am Main, Germany}
\address{$^2$ Department of Physics, St. Petersburg State Maritime Technical University, Leninskii prospect 101, St. Petersburg 198262, Russia}
\ead{lo@fias.uni-frankfurt.de}

%===============================================================================
%ABSTRACT
\begin{abstract}
The present work is a generalisation of the dynamical screening factor presented in \cite{LoKorol07} to consider an atom located at an arbitrary position within the fullerene.
A more elaborated investigation into the case where the atom is located at the centre is performed and compared with quantum mechanical calculations for dynamical screening factor of Ar@C$_{60}$ \cite{MadjetChakraborty07} and Mg@C$_{60}$ \cite{ChakrabortyMadjet08}.
The $\pi$ and $\sigma$ plasmons of the fullerene are accounted for in a modified screening factor to improve correspondence with the quantum calculations.
The spatial dependence of the screening factor was explored with Ar@C$_{60}$ and Ar@C$_{240}$ and found to depend significantly on the radial distance of the atom from the centre of the fullerene.
A spatial averaging of the screening factor is presented.
\end{abstract}

%\submitto{\jpb}
%================================================================================
%SECTION: INTRODUCTION
\section{Introduction}

Particles may be confined by a wide range of methods.
These include trapping by optical means, confinement within materials, e.g. as dopants in solids, or within hollow structures such as fullerenes and nanotubes.
As the ability to insert atoms or molecules into these carbon cages
increased, so too has the interest in these endohedral systems grown \cite{Sliwa96,Shinohara00}.
The possible applications of this novel form of matter ranges from forming the register of a quantum computer \cite{Harniet02,BenjaminArdavan06,Tomanek05} to enhancing imaging techniques \cite{FatourosCorwin06}.

Understanding the properties of such systems is therefore of much interest. One of the means of achieving this is by spectroscopic means.
If the confining effect of the cage on the optical properties of the endohedral species can be understood, this may provide a means of predicting the properties of the whole system.

The fullerene cage screens the endohedral atom from external electromagnetic fields. Recent theoretical studies \cite{ConneradeSolovyov05,LoKorol07,LoKorolISACC08}, using semi-classical methods,  show that this effect is strongly dependent on the photon energy.
It was found that in the vicinity of the plasmon resonances of the fullerene, the photoabsorption of the confined atom is strongly enhanced from that of the free atom.
A dynamical screening factor may be defined to relate the two photoabsorption cross sections.

There are other considerations on how the photoabsorption or photoionization of the atom is modified by the confining fullerene cage.
Confinement resonances, as discussed in \cite{Baltenkov99,ConneradeDolmatov00}, consider the interference caused by the photoelectron reflected by the fullerene's inner and outer surfaces.
The modification of the cross section owing to the interference between the confinement resonances and intra-doublet atomic resonances are investigated in \cite{AmusiaChernysheva02}, with reference to the experimental work of \cite{KivimakiHergenhahn00}.
This was elaborated upon in \cite{AmusiaBaltenkov07,AmusiaBaltenkov08}.
Reference \cite{DolmatovManson08} discusses the effect of multi-electron correlation on the confinement resonances.
Also, the role of multi-walled fullerene cages is studied in \cite{DolmatovBrewer08}.

In our previous work on the dynamical screening by the fullerene \cite{ConneradeSolovyov05,LoKorol07}, the endohedral atom is located at the centre of the fullerene.
However, interactions with the fullerene, thermal vibrations, electron transfer, etc, leads to the atom being not necessarily fixed at any one position and that the dynamical screening factor should be considered over all possible positions of the confined species.
In \cite{LoKorol07}, the dynamical screening factor was presented for the limiting case where the mutual interaction between the polarized atom and the polarized fullerene cage was neglected.
Within this approximation, the screening factor is independent of the position of the endohedral atom.
In the present paper, a more detailed analysis of the case of the centrally positioned atom is performed.
Also, a comparison of this model with results from time dependant local density approximation calculations on Ar@C$_{60}$ \cite{MadjetChakraborty07} and Mg@C$_{60}$ \cite{ChakrabortyMadjet08} is elaborated upon.
The $\pi$ and $\sigma$ plasmons of the fullerene are accounted for in a modified screening factor to improve correspondence with the quantum calculations and, on this basis, to achieve a better understanding of the physics of the dynamical screening effect.
The spatial dependence of the dynamical screening factor is investigated for two case studies: Ar@C$_{60}$ and Ar@C$_{240}$.
A method for spatial averaging and its results are presented.

The atomic system of units, $e = m = \hbar = 1$ is used throughout the paper.

%===============================================================================
%SECTION: THEORETICAL FRAMEWORK
\section{Theoretical Framework}
\label{sec:theory}

%===== intro =====
When the endohedral system is exposed to an external electromagnetic  field, the fullerene, being dynamically polarized, will screen the confined atom.
Depending on the frequency of the light, the field at the atom may be stronger or weaker.
Therefore, the photoabsorption cross section of the confined atom can differ from that of the free atom in the same external field.
A dynamical screening factor $\cF \equiv \cF(\omega)$ may be defined to relate the two:
\begin{equation}
\label{eq:definition}
\cF
=
\frac{\sigma_{\rm{conf}}}{\sigma_{\rm{free}}}.
\end{equation}
%

%===== about model ======
A semi-classical approach is taken here to study this effect.
The fullerene is modelled as a dielectric shell \cite{LambinLucas92,OestlingApell93a,OestlingApell96,AndersenBonderup00,Sihvola06} and classical electrostatic methods are used.
As in \cite{LoKorol07,LoKorolISACC08}, the fullerene is treated as a spherical shell of some thickness $\Delta R$ with an inner and outer radius of $R_1$ and $R_2$ respectively.
\begin{figure}
\centering
\includegraphics[width=0.65\textwidth]{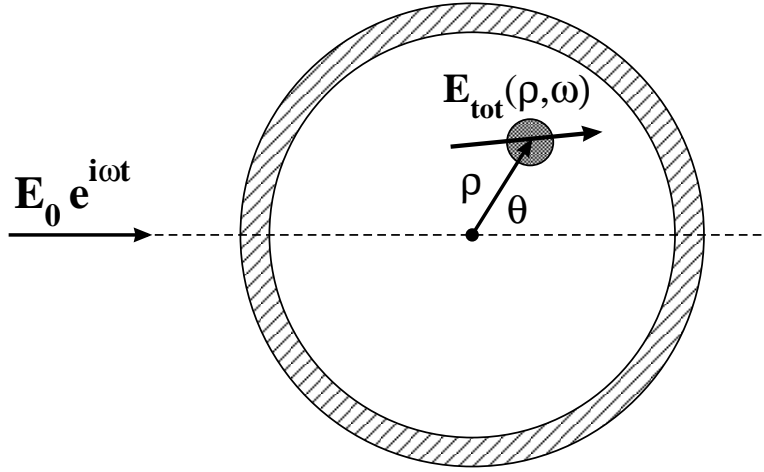}
\caption{The position of the point-like atom, indicated by the filled circle, is confined at $\bfrho$ inside the fullerene.
The angle between $\bfrho$ and the external field $\bfE_0$ is denoted $\theta$.
The field at the position of the fullerene is given by $\bfE_{\mathrm{tot}}(\bfrho,\omega)$.}
\label{fig:noncentral_diagram}
\end{figure}
%
%===== DRUDE =====
The dielectric function of the fullerene is $\epsilon \equiv \epsilon(\omega)$.
Given the strong delocalisation of valence electrons of the fullerene, they are treated as a free electron gas.
The dielectric function of the fullerene $\epsilon(\omega)$ may be expressed via the Drude model \cite{BohrenHuffman}:
\begin{equation}
\label{eq:drude}
\epsilon(\omega) = 1 - \frac{\omega_{\rm{p}}^2}{\omega^2}.
\end{equation}
The plasma frequency $\omega_{\rm{p}}$ is:
\begin{equation}
\label{eq:wp}
\omega^2_{\rm{p}} = \frac{4\pi N}{V},
\end{equation}
where N is the number of valence electrons and $V = 4\pi (R_2^3 -R_1^3) / 3$ is the volume of the fullerene shell.

%===== point like atom ====

The point-like endohedral atom is characterised by the dynamic dipole polarizability $\alpha(\omega)$ and can be located at an arbitrary position $\bfrho$ within the fullerene, see figure \ref{fig:noncentral_diagram}.

We consider the case when the endohedral system is in a uniform and monochromatic external field, whose strength is given by $\bfE_\ext(t) = \bfE_0 \re^{\ri \omega t}$, where $\bfE_0$ is a real amplitude.
\\

The photoabsorption cross section of an object is proportional to its photoabsorption rate, $Q$, which can be calculated from the relation \cite{Landau8,AndersenBonderup00}:
\begin{equation}
\label{eq:Q}
Q = \frac{\omega}{2} \, \rm{Im} \int \bfE^*(\bfr) \, d\bfD(\bfr).
\end{equation}
Here $\bfE(\bfr)$ and $\rd\bfD(\bfr)$ are the electric field and the elementary dipole moment at the position $\bfr$.
The integration is carried out over the volume of the considered system.
For a free point-like atom the photoabsorption rate becomes
\begin{equation}
Q_{\mathrm{free}}(\omega)
=
\frac{\omega}{2} \, \rm{Im} \left( \bfE_0^* \bfD \right)
=
\frac{\omega}{2} \, E_0^2 \,\rm{Im} \left( \alpha(\omega) \right),
\label{eq:Q_free}
\end{equation}
with $\bfD = \bfD(\omega) = \alpha(\omega) \bfE_0$ being the induced dipole moment of the atom.
Similarly, the photoabsorption rate of the confined atom is
\begin{equation}
Q_{\mathrm{conf}}(\bfrho,\omega)
=
\frac{\omega}{2} \, \rm{Im}
\left( \bfE_{\tot}^*(\bfrho,\omega) \bfD_{\tot}(\bfrho,\omega) \right)
=
\frac{\omega}{2} \, E_{\tot}^2(\bfrho,\omega) \,\rm{Im} \left( \alpha(\omega) \right),
\label{eq:Q_confined}
\end{equation}
where $\bfE_{\tot}(\bfrho,\omega)$ is the total electric field at the position $\bfrho$ and $\bfD_{\tot}(\bfrho,\omega)$ is the atomic dipole moment, which may depend on the position of the atom and the frequency of the field.
Thus, the dynamical screening factor becomes
\begin{equation}
\label{eq:cFdefinition}
\cF(\bfrho,\omega)
=
\frac{Q_{\mathrm{conf}}}{Q_{\mathrm{free}}}
=
\frac{E_{\tot}^2(\bfrho,\omega)} {E_0^2}.
\end{equation}

%===== screening factor from Korol =====
In order to calculate the dynamical screening factor, it is necessary to calculate the total field at the position of the atom.
The total field deviates from the external field: $\bfE_{\tot} = \bfE_0 + \Delta\bfE(\bfrho,\omega)$.
Here, the additional term $\Delta\bfE(\bfrho,\omega)$ appears as a result of the fullerene's polarization.
The latter is due to the action of both the external field {\it and} the field of the dipole induced at the atom.
If the atomic polarizability $\alpha(\omega)$ is small, the inverse effect of the polarized atom on the shell can be ignored.
In this case, the term $\Delta\bfE(\bfrho,\omega)$ is solely due to the external field so that it does not depend on the position vector $\bfrho$.
This means that the total field and, consequently, the dynamic screening factor are not sensitive to the atom's position inside the fullerene cage.
Such an approach was adopted in previous studies \cite{LoKorol07}.

However, the polarizability of the endohedral atom may be large at certain photon energies, primarily in the vicinities of giant resonances in the atomic photoabsorption cross section.
In these energy regions the inverse effect of the polarized atom on the cage may become significant.
As a result, the induced moments of the atom and the cage become correlated and acquire the dependence on $\bfrho$.
To calculate the additional field $\Delta\bfE(\bfrho,\omega)$ one has to carry out a self-consistent consideration which takes into account the mutual polarization of the atom and the fullerene.

The self-consistent method is based on the iterative procedure developed in \cite{LoKorol07}.
For the sake of clarity, let us outline the main points of this procedure.

To begin with, one makes use of the following two auxiliary problems.

The well-known first problem (see, e.g. \cite{BohrenHuffman}) concerns the field in the interior of a dielectric spherical shell exposed to the uniform external field $\bfE_0$.
The field in question can be written as $(1-z)\bfE_0$, where the quantity $z$ depends on the dielectric function $\epsilon$ of the shell, and on its inner and outer radii.
In the case of a monochromatic field $z$ also acquires the dependence on $\omega$.

The second problem is the determination of the electric field $\bfE$ due to the polarization of the shell by a point-like dipole $\bfD$ located at the position $\bfrho$ inside the shell.
Using standard methods of electrostatics, one can demonstrate that at the position $\bfrho$ this field can be represented as follows:
\begin{equation}
\bfE(\bfrho)
=
\hat{s}(\bfrho) \bfD
\equiv
-\frac{1}{R_1^3}
\left[s_1(\rho) \bfD + s_2(\rho) \left( \bfD \bfn \right) \bfn \right].
\label{eq:vdefn}
\end{equation}
Here $\bfn$ is the unit vector in the direction of $\bfrho$ and $s_{1,2}(\rho)$ are the scalar functions, which depend, apart from $\rho$, on $\epsilon$, $R_1$ and $R_2$.

%===== the iteration =====
Then, the iterative procedure is as follows.
At first, the external field polarizes the fullerene, giving rise to an additional field in the interior.
These two fields combined give the field inside the fullerene $\bfE_1 = (1-z) \bfE_0$, which induces a dipole moment in the atom $\bfd_1 = \alpha(\omega) \bfE_1$.
In the next iterative step, the dipole moment of the atom acts on the fullerene (problem 2), bringing about another field $\bfE_2 = \hat{s}(\bfrho) \, \bfd_1$ at the atom.
An additional dipole moment is induced in the atom: $\bfd_2 = \alpha(\omega) \, \bfE_2$.
For the $i^\mathrm{th}$ iteration: $\bfE_i = \hat{s}(\bfrho) \, \bfd_{i-1}$ and $\bfd_i = \alpha(\omega) \, \bfE_i$.
The total field is therefore:
\begin{eqnarray}
\bfE_\tot(\bfrho) = \bfE_1 + \bfE_2 + \bfE_3 + \dots
\label{eq:Eiteration}
\end{eqnarray}
and the total dipole moment of the atom is $\bfD_\tot(\bfrho) = \alpha(\omega)\bfE_\tot(\bfrho)$.\\

%===== PLASMON RESONANCE APPROXIMATION =====
Evaluating the series from (\ref{eq:Eiteration}), one derives the following expression for the total field at the position of the atom:
\begin{equation}
\label{eq:noncentral_totalfield}
\bfE_{\mathrm{tot}}(\bfrho,\omega)
=
\frac{1-z(\omega)}{ 1 + s_1\,\alpha(\omega)/R_1^{3}}
\left[
\bfE_{0}
-
\frac{s_2\,\alpha(\omega)/R_1^{3}}
{ 1 + (s_1+s_2)\,\alpha(\omega)/R_1^{3}}
 \,\Bigl(\bfn\bfE_0\Bigr)\,\bfn
\right].
\label{eq:Etotal}
\end{equation}
The quantity $z(\omega)$ characterizes the dynamical response of the fullerene to the external field and is defined as:
\begin{equation}
z(\omega)
=
\frac{2(N_{l2}-N_{l1})}{R_2^3}
\left(
\frac{1}{\omega^2 - \omega_{1l}^2 + \ri  \Gamma_{1l} \omega}
-
\frac{1}{\omega^2 - \omega_{2l}^2 + \ri \Gamma_{2l} \omega} 
\right)\bigg|_{l=1}.
\label{eq:z}
\end{equation}
The surface plasmon frequencies are denoted by $\omega_{1l}$ and $\omega_{2l}$ for a given multipole moment $l$.
These pairs of eigenmodes arise from the fullerene having a finite thickness and therefore two surface charge densities.
The eigenfrequency $\omega_{1l}$ characterises the symmetric mode, in which the two surface charge densities oscillate in phase, whereas $\omega_{2l}$ characterises the anti-symmetric mode, in which the oscillations are in anti-phase \cite{LoKorol07}.
These quantities are defined as:
\begin{eqnarray}
\cases{
\displaystyle{
\omega_{1l}^2
=
\frac{\omega_p^2}{2(2l+1)} \left(2l+1-p_l\right)
}
\\
\displaystyle{
\omega_{2l}^2 
=
\frac{\omega_p^2}{2(2l+1)}
\left(2l+1+p_l\right)
}}
\qquad
\cases{
\displaystyle{
N_{l1} 
= N\,\frac{p_l+1}{2p_l}
}
\\
\displaystyle{
N_{l2} 
= N \,\frac{p_l-1}{2p_l}
}}
\label{eq:plasmon_freq}
\end{eqnarray}
where
\begin{equation}
p_l = \sqrt{1 + 4l(l+1)\xi^{2l+1}},
\end{equation}
with $\xi = R_1/R_2 \leq 1$ being the ratio of the fullerene's inner and outer radii.
The corresponding widths and oscillator strengths of these modes are denoted by $(\Gamma_{jl}, N_{jl})_{j=1,2}$.
Within the framework of our approach, the widths of these surface plasmon modes are not computed but are treated as parameters.

In (\ref{eq:Etotal}), both $s_{1}$ and $s_{2}$ are related to the multipole expansion of the electric field in the situation, where a point-like dipole is positioned at $\bfrho$ inside the fullerene:
$s_{1}$ is due to the field component parallel to the point like-dipole while $s_{2}$ is due to the component perpendicular to the dipole.
Their explicit forms are:
\begin{equation}
\left\{
\begin{array}{c}
s_1\\
s_2
\end{array}
\right\} 
=
\frac{1}{2}
\sum_{l=1}^{\infty}
\left\{
\begin{array}{c}
l(l+1)^2
\\
l(l^2-1)
\end{array}
\right\} 
\left(\frac{\rho}{R_1}\right)^{2l-2}
\beta_{l}(\omega)\,
\frac{1 - \xi^{2l+1}}{1-\xi^3}.
\end{equation}

The quantity $\beta_l(\omega)$ characterizes the dynamical response of the shell to an electric field of the multipolarity $l$ and the frequency $\omega$:
\begin{equation}
\beta_{l}(\omega)
=
\frac{1}{R_2^3} \, \frac{3}{2l+1}
\left(
\frac{N_{l2}}{\omega^2 - \omega_{1l}^2  + \rm{i} \Gamma_{1l} \omega}
+
\frac{N_{l1}}{\omega^2 - \omega_{2l}^2  + \rm{i} \Gamma_{2l} \omega}
\right).
\end{equation}
%
%\\
%===== SCREENING FACTOR =====
Combining equations (\ref{eq:cFdefinition}) and (\ref{eq:noncentral_totalfield}), one arrives at the following expression for the dynamical screening factor:
\begin{eqnarray}
\fl
\cF(\bfrho,\omega)
=
\frac{\Bigl|1-z(\omega)\Bigr|^2}{\left|1 + s_1\, \alpha(\omega)/R_1^3\right|^2}
\label{eq:factor}
\\
\times
\left(
1
+
\left[
\left|\frac{s_2\,\alpha(\omega)/R_1^3}{1+ (s_1+s_2)\alpha(\omega)/R_1^3}
\right|^2
-
2\mathrm{Re}\frac{s_2 \,\alpha(\omega)/R_1^3}{1 + (s_1+s_2)\alpha(\omega)/R_1^3}
\right]
 \cos^2\theta
\right),
\nonumber
\end{eqnarray}
where $\theta$ is the angle between $\bfrho$ and $\bfE_0$ (see figure \ref{fig:noncentral_diagram}).\\
%

%===== CENTRAL CASE =====
%\subsection{The central case, $\bfrho = 0$}
Let us apply (\ref{eq:factor}) to the special case where the atom is located at the central position.
Because of the symmetry of this system, the multipole expansion of the electric field inside the fullerene is limited to the dipole term.
Thus, the terms $s_1$ and $s_2$ reduce to:
\begin{equation}
s_1(0,\omega) = 2\beta_{1}(\omega),
\qquad
s_2(0,\omega) = 0,
\end{equation}
and the total field at the atom simplifies to
\begin{equation}
\bfE_{\mathrm{tot}}(0,\omega)
=
\frac{1-z(\omega) }
{1 + 2\beta_{1}(\omega)\,\alpha(\omega)/R_1^{3}}
\,\bfE_{0}.
\end{equation}
The dynamical screening factor becomes
\begin{equation}
\cF(0,\omega)
=
{|1-z(\omega)|^2 
\over \left|1 {+} 
2\beta_{1}(\omega)\alpha(\omega)/{R_1^{3}}\right|^2}.
\label{eq:factor_central}
\end{equation}
In the limiting case, where the interaction between the polarized atom and the polarized fullerene shell is negligible, the denominator of equation (\ref{eq:factor_central}) goes to 1, and one obtains
\begin{equation}
\cF(0,\omega)
\approx
|1-z(\omega)|^2.
\label{eq:factor_central_nopol}
\end{equation}
This result for $\cF$ initially derived in \cite{LoKorol07}, is independent of the nature of the endohedral atom.

%=====================RESULTS!!!================================
\section{Numerical Results}

The dynamical screening factor is sensitive to a number of quantities such as the size and thickness of the fullerene cage, the type of endohedral atom and its position within the fullerene.
To illustrate these dependencies, we begin by considering the case where the atom is located at the centre of the fullerene.
Whenever possible, we compare our results with those presented in \cite{MadjetChakraborty07,ChakrabortyMadjet08}, where the problem of dynamical screening was considered within the framework of the time-dependent LDA (TDLDA).

%===== the central case =====
\subsection{Centrally positioned atom}
\label{sec:results_central}
To investigate the impact of the type of endohedral atom on the dynamical screening factor, two endohedral systems are considered: Ar@C$_{60}$ and Mg@C$_{60}$.
In this section the analysis is carried out for the central position of the atom, i.e. $\bfrho=0$ (see figure \ref{fig:noncentral_diagram}).

\begin{figure}
\centering
\includegraphics[width=0.7\textwidth]{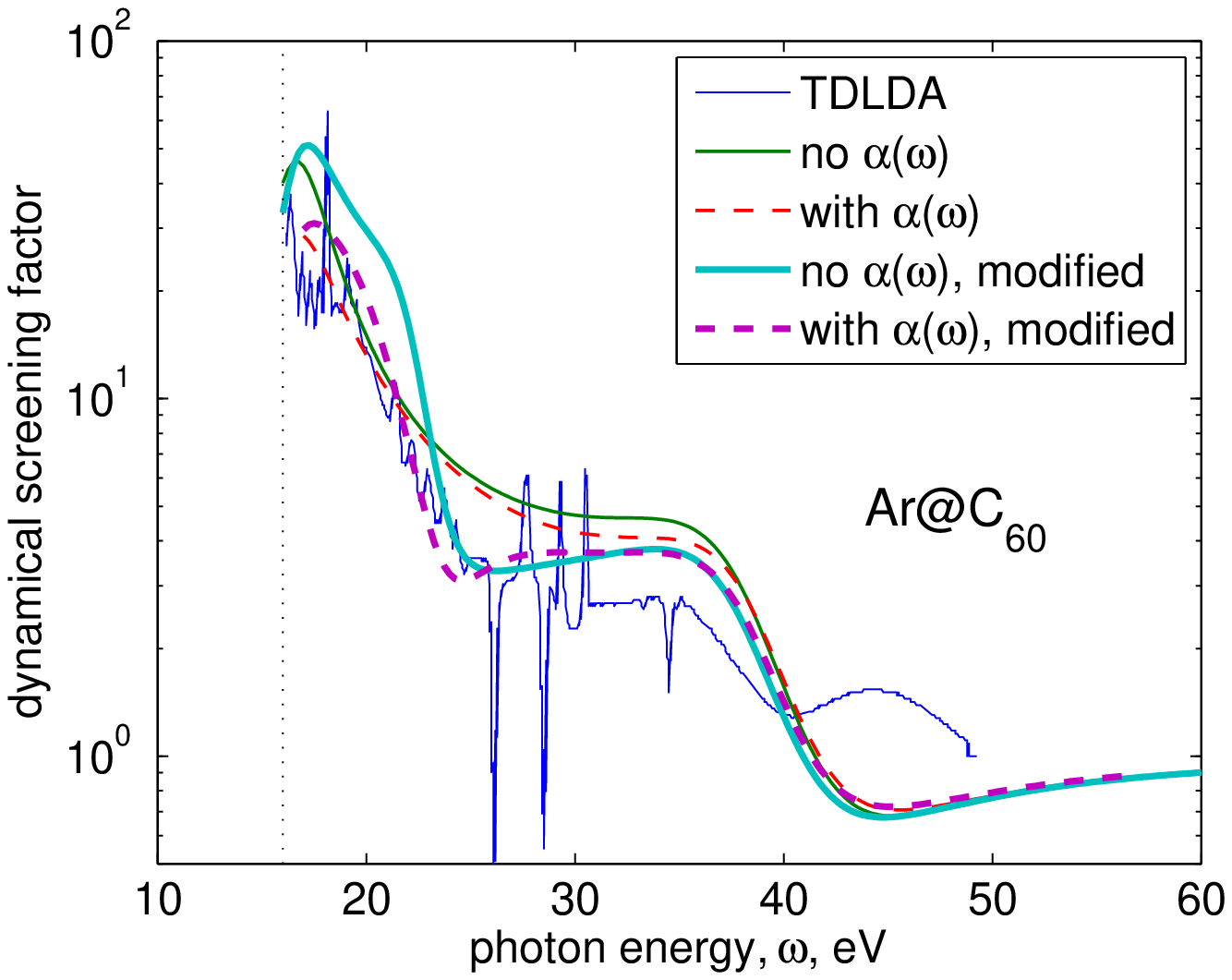}\\
\includegraphics[width=0.7\textwidth]{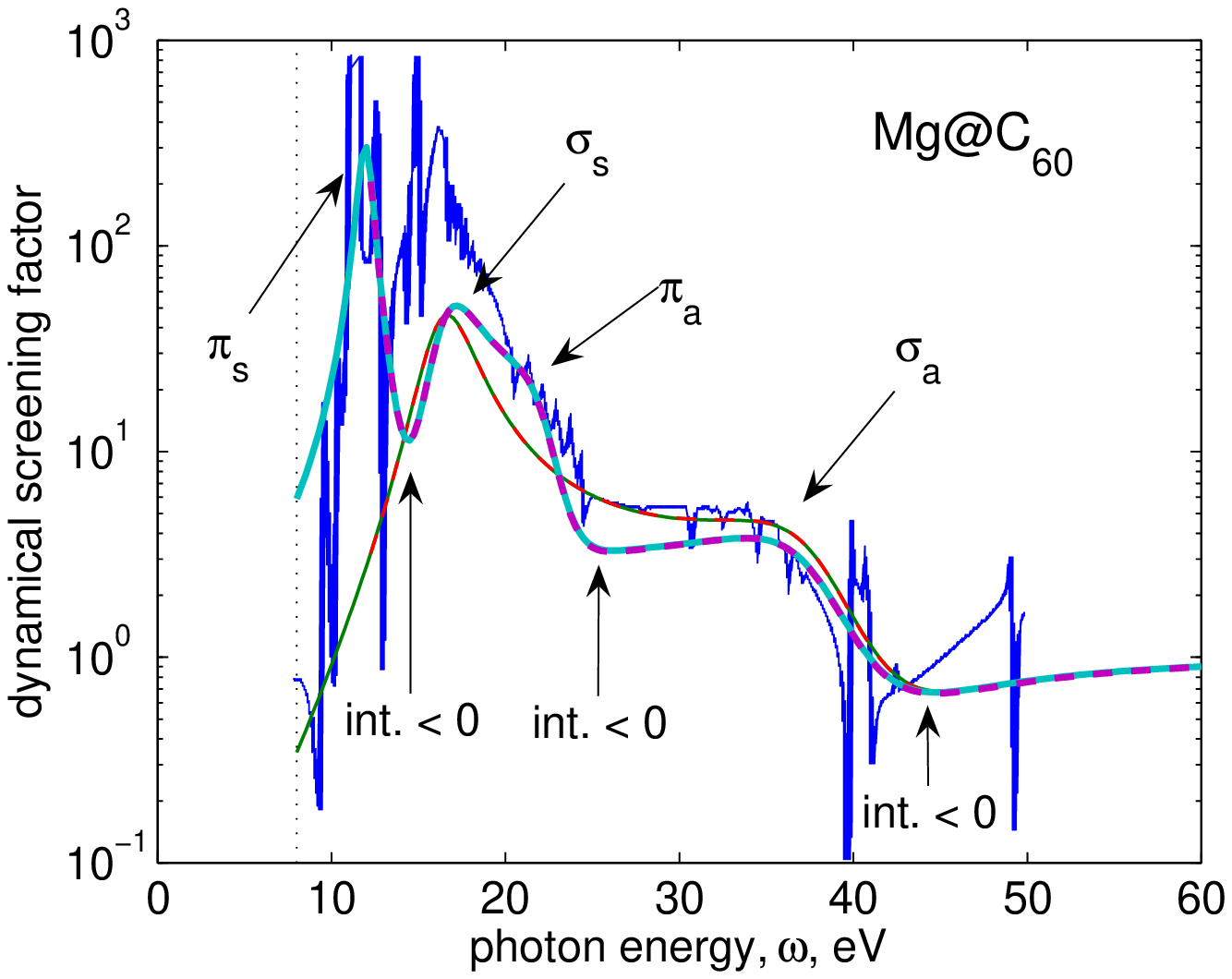}
\caption{
The dynamical screening factor for the centrally positioned endohedral atom: Ar@$C_{60}$ (top) and Mg@$C_{60}$ (bottom).
The common legend is given in the top panel.
The dynamical screening factor was calculated for two cases: from (\ref{eq:factor_central_nopol}), which neglects the atomic feedback, labelled as `no $\alpha(\omega)$', and from (\ref{eq:factor_central}), with accounting for this feedback, labelled as `with $\alpha(\omega)$', represented respectively by the solid and the dashed lines.
These are compared with TDLDA calculations (thin solid line) from \cite{MadjetChakraborty07} and \cite{ChakrabortyMadjet08}.
The ionization potential of the endohedral atom is indicated by the dotted line.
The thick solid and dashed lines represent the factor calculated within the modified model, which includes the contributions of the $\sigma$ and $\pi$ plasmons, see (\ref{eq:factor_pi_nopol}) and (\ref{eq:factor_pi_pol}).
The symmetric (s) and antisymmetric (a) modes of the $\pi$ and the $\sigma$ plasmons are indicated by arrows on the plot for Mg@C$_{60}$.
Regions of particularly strong interference (`int.'), all negative, between the $\pi$ and the $\sigma$ plasmons are also indicated.
}
\label{fig:central}
\end{figure}
\begin{figure}
\centering
\includegraphics[width=0.45\textwidth]{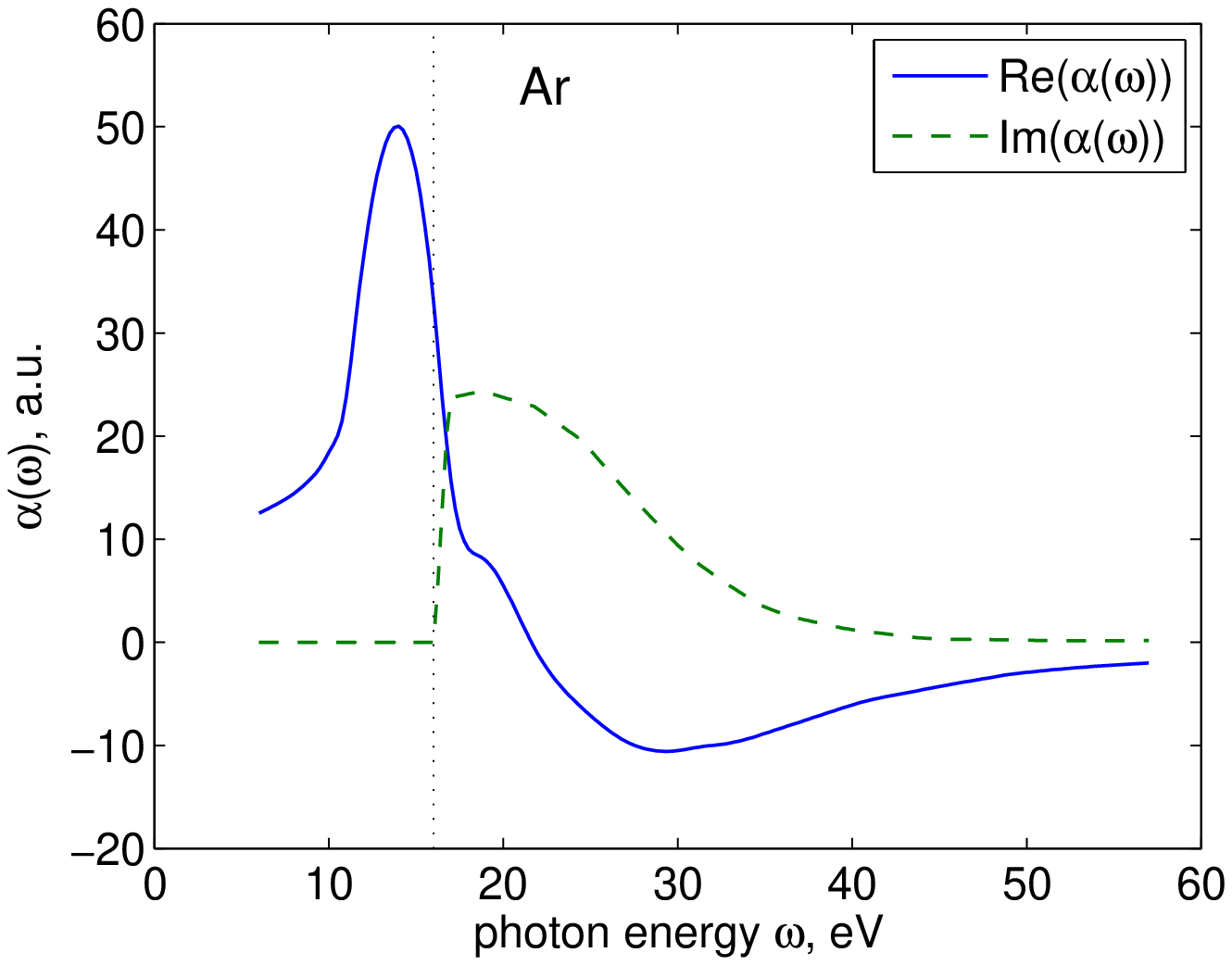}
\includegraphics[width=0.45\textwidth]{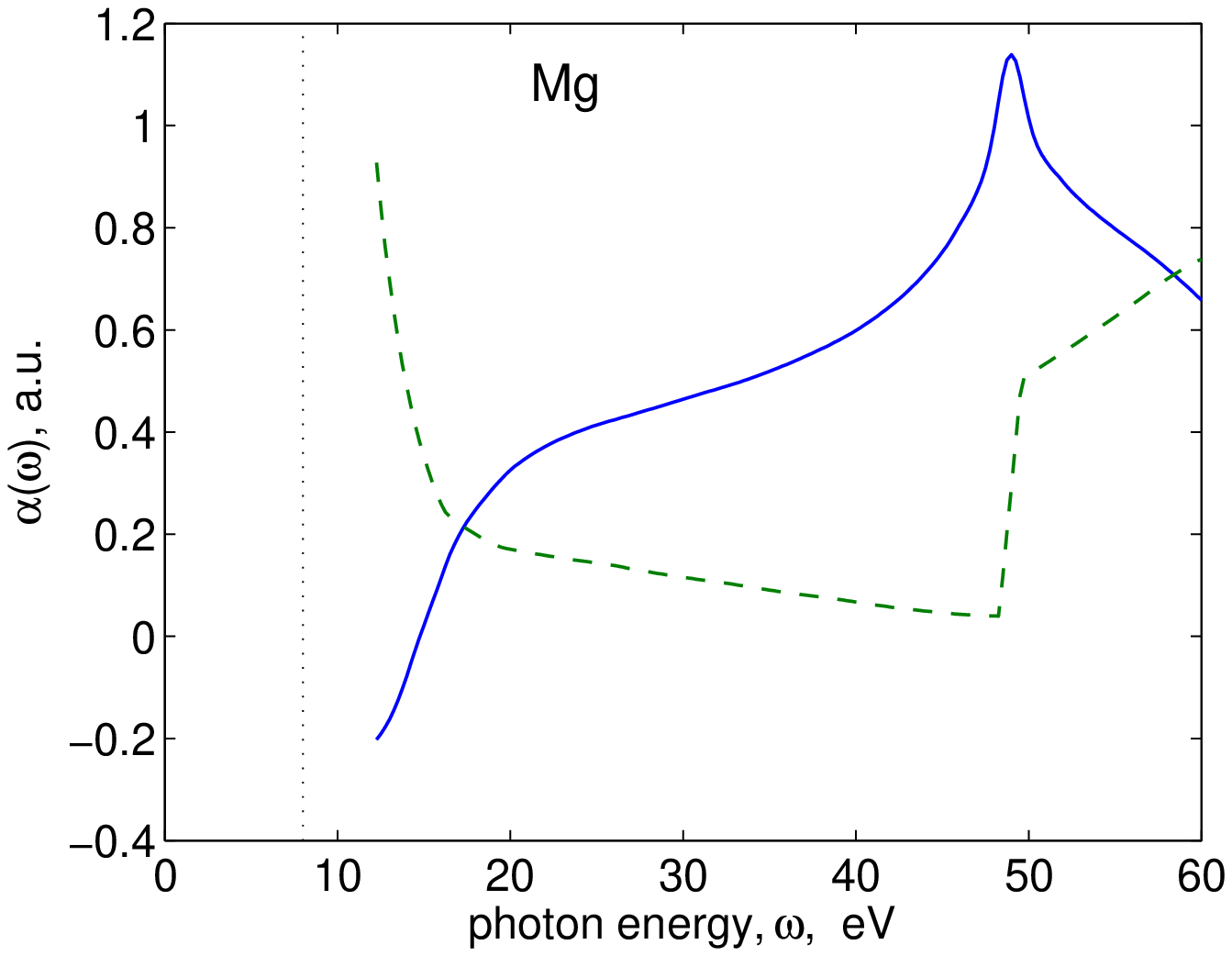}
\caption{Left: the dynamic dipole polarizability of argon, calculated within the random phase approximation with exchange (RPAE).
The common legend is given in the right panel.
Right: the dynamic dipole polarizability of magnesium, based on the scattering factors of \cite{Henke93}.
The real was calculated from the imaginary part of the polarizability via the dispersion relation.
In both plots, the solid blue line is the real part and the dashed green line is the imaginary part.
The first ionization threshold of the atom is indicated by the dotted line.}
\label{fig:pol}
\end{figure}

% exact case
The dynamical screening factors (\ref{eq:factor_central}) for these two systems are shown in figure \ref{fig:central} as smooth solid lines of medium thickness.
For each system, it begins above the ionization threshold, $I$, of the endohedral atom, which is 16 eV for argon and 8 eV for magnesium. 
For $\omega<I$ the calculation of the screening factor according to the definition (\ref{eq:definition}) is meaningless, since the atomic photoionization cross section is identically equal to zero.
% limiting case

For each atom, the curve is compared with the limiting case where the interaction between the atom and fullerene is neglected (\ref{eq:factor_central_nopol}).
The corresponding universal curve, independent of the type of confined atom, is presented in both graphs as a smooth dashed line of medium thickness.
% comparison
It is seen that the screening factor can depend strongly on the type of endohedral atom.

For argon, there are noticeable differences between the two cases, whereas for magnesium the two curves lie almost on top of each other.
This can be understood by considering the magnitudes of the dynamic dipole polarizabilities of the atoms, given in figure \ref{fig:pol}.
The polarizability of argon is large in this energy range, so the inverse effect of the polarized atom on the fullerene is significant.

%===== parameters =====
In these calculations, the following parameters were used:
the two dipole surface plasmon resonances of the fullerene were set at 16.5 and 38.0 eV respectively, with their widths being 3.5 and 9.0 eV.
Additionally, the mean radius and the thickness of the fullerene were set to 3.54 \AA\,  and 1.5 \AA, respectively.
These parameters are the same as those used by Chakraborty and co-workers in the TDLDA calculations of the dynamical screening factor for Ar@C$_{60}$ \cite{MadjetChakraborty07} and Mg@C$_{60}$ \cite{ChakrabortyMadjet08}.
Their results are shown in figure \ref{fig:central} as thin spiky solid lines.

%===== comparision with manson =====
In the case of Ar@C$_{60}$, there is a good agreement between the model results and the TDLDA calculation.
The inclusion of the interaction between the polarized atom and fullerene into the dynamical screening factor improves the correspondence.
However, in the case of magnesium, there is only a qualitative agreement in the higher energy range.
At about 10 eV, there is a strong narrow peak in the TDLDA calculation that cannot be described by our model.
There is also a large difference about the symmetric plasmon mode at 16.5 eV.\\

%===== pi plasmon =====
The following is a possible explanation for the large discrepancies for Mg@C$_{60}$.
There are two types of valence electrons of the fullerene: the $\sigma$ and the $\pi$ electrons \cite{FullereneAtlas}.
The $\pi$ electron is more strongly bound to the carbon core whereas the $\sigma$ electron is more weakly bound.
From this arises two types of surface plasmons: the $\pi$ plasmon and the $\sigma$ plasmon (see, e.g.  \cite{OestlingApell93a,Solovyov05}).
The $\sigma$ plasmon involves mainly the more loosely bound electrons and is located at about 20 eV while the $\pi$ plasmon involves only the strongly bound ones, resulting in a lower plasmon energy of around 10 eV \cite{OestlingApell93a,IvanovKashenock01,AndersenBonderup00}.
 
Let us introduce the two types of plasmons into the dynamical screening factor.
To do so, one models the fullerene as a system of two co-centric spherical shells.
The valence electrons of the fullerene are distributed into these shells, so that one shell contains those, which participate in the $\pi$ plasmon, and the other contains the electrons from the $\sigma$ plasmon.
In reality, the plasmons will influence each other's resonance energy.
We take the simplest approach, in which this effect is ignored.

The finite thickness of the shells leads to the splitting of each plasmon into a symmetric and an antisymmetric mode.
As observed in \cite{ReinkoesterKorica04,ScullyEmmons05} and seen in other theoretical works on these plasmon modes of the fullerene (e.g. \cite{OestlingApell96,KorolSolovyov07}), the antisymmetric mode does not generally manifest itself as strongly as the other mode.

The dynamical screening factor of such a system is calculated using the iterative scheme described in section \ref{sec:theory}.
The only modification to the scheme is the necessity of accounting for the two shells, which contribute to the total electric field at the centre.
Thus, at each iterative step, there is a contribution to the electric field at the atom due to the $\pi$ shell and to the $\sigma$ shell. 

With this modification, the dynamical screening for the centrally positioned atom becomes (cf. (\ref{eq:factor_central})):
\begin{equation}
\cF(\omega)
=
\left|
\frac{1 - z_\pi(\omega) - z_{\sigma}(\omega)}
{1+2\alpha(\omega) \left( \beta_{1,\pi}/R_{1,\pi}^3 + \beta_{1,\sigma}/R_{1,\sigma}^3  \right) }
\right|^2.
\label{eq:factor_pi_pol}
\end{equation}
The properties of the endohedral atom are encoded in its dynamic dipole polarizability $\alpha(\omega)$.
For the case where the polarizability of the confined atom is negligible, this is simply
\begin{equation}
\cF(\omega)
\approx
\left|
1 - z_\pi(\omega) - z_{\sigma}(\omega)
\right|^2.
\label{eq:factor_pi_nopol}
\end{equation}

The modified dynamical screening factor is shown in figure \ref{fig:central} as thick lines.
The solid line is the dynamical screening factor for the limiting case and the dashed line accounts for the interaction between the polarized atom and the polarized fullerene.

For Mg@C$_{60}$, the pair of curves lie on top of each other, owing to the small dynamic dipole polarizability of magnesium in this energy range (see figure \ref{fig:pol}, right panel).
There is an improvement in the correspondence between our model (\ref{eq:factor_pi_pol}) and the TDLDA calculations of \cite{ChakrabortyMadjet08}.
We have now reproduced the sharp peak near 10 eV that was missing from the original model.
Within the modified model, this feature is due to the symmetric mode of the $\pi$
plasmon.
Also reproduced is the additional feature near 20 eV arising from the antisymmetric mode.
The positions of the 4 plasmon modes of the modified model are all indicated in figure \ref{fig:central} in the bottom panel.
As most clearly demonstrated in (\ref{eq:factor_pi_nopol}), the screening factor consists of direct squares and cross multiplication terms, which raise or lower it depending on the photon energy.
These cross terms, or interference terms, are particularly significant in 3 regions indicated in the lower panel of figure \ref{fig:central} by arrows labelled `int.'.
In these regions, the interference is strongly negative, resulting in well defined minima in the profile of the screening factor.
There is still a quantitative difference near the symmetric mode of the $\sigma$ plasmon.
It is possible to reduce this difference by increasing $N_{\sigma}$, shifting $\omega_{\pi,2}$ towards $\omega_{\sigma,1}$, and reducing the corresponding widths.

For Ar@C$_{60}$, the modification of the screening factor due to the $\pi$ and the $\sigma$ plasmons has resulted in an even better correspondence with the TDLDA calculation from \cite{MadjetChakraborty07}.

In these calculations, $N_{\pi}$ and $N_{\sigma}$ were fixed to 80 and 160 respectively.
The inner and outer radii and thicknesses of the two shells were adjusted to fit the energies of the plasmons observed in the TDLDA calculation for Mg@C$_{60}$.
The parameters are given in table \ref{tab:pi_parameters} along with the energies $\omega_{1,2}$ and widths $\Gamma_{1,2}$ of the symmetric and antisymmetric plasmon modes of each shell.
\begin{table}[t]
\caption{
Given below are the parameters used to calculate the modified dynamical screening factor
(\ref{eq:factor_pi_pol}) for Ar@C$_{60}$ and Mg@C$_{60}$.
The fullerene is modelled as two spherical shells - the $\pi$ shell and the $\sigma$ shell.
Their radii are given in Angstroms.
Also given (in eV), are the resulting plasmon energies and the widths that are assigned to the plasmons.}
\label{tab:pi_parameters}
\lineup
\begin{indented}
\item[]
\begin{tabular}{@{}llllllll}
\br
& $N$ & $R_1$ & $R_2$ & $\omega_1$ & $\omega_2$ & $\Gamma_1$ & $\Gamma_2$ \\
\mr
$\pi$ shell & \080 & 2.25 & 3.75 & 12.0 & 22.2 & 1 & 4\\
$\sigma$ shell & 160 & 2.7 & 3.7 & 16.6 & 37.7 & 3.5 & 9\\
\br
\end{tabular}
\end{indented}

\end{table}

%===== the general case =====
\subsection{Arbitrarily positioned atom}

In reality, the endohedral atom is not fixed at the centre of the fullerene.
Thermal vibrations, van der Waals interaction with the fullerene, and other effects, such as electron transfer, can cause the atom to be preferentially displaced from the centre.

\begin{figure}[t]
\centering
\includegraphics[width=0.45\textwidth]{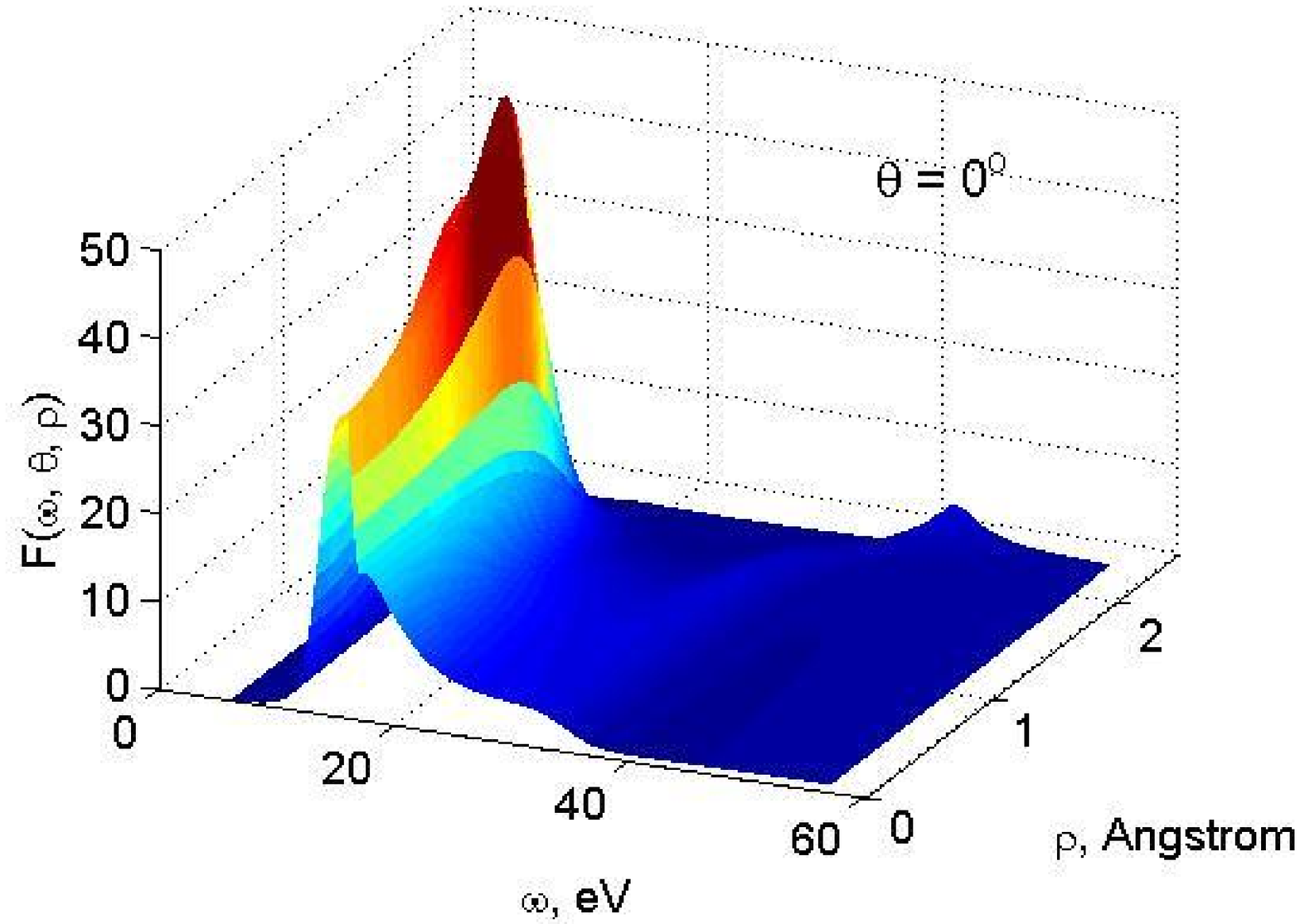}
\includegraphics[width=0.45\textwidth]{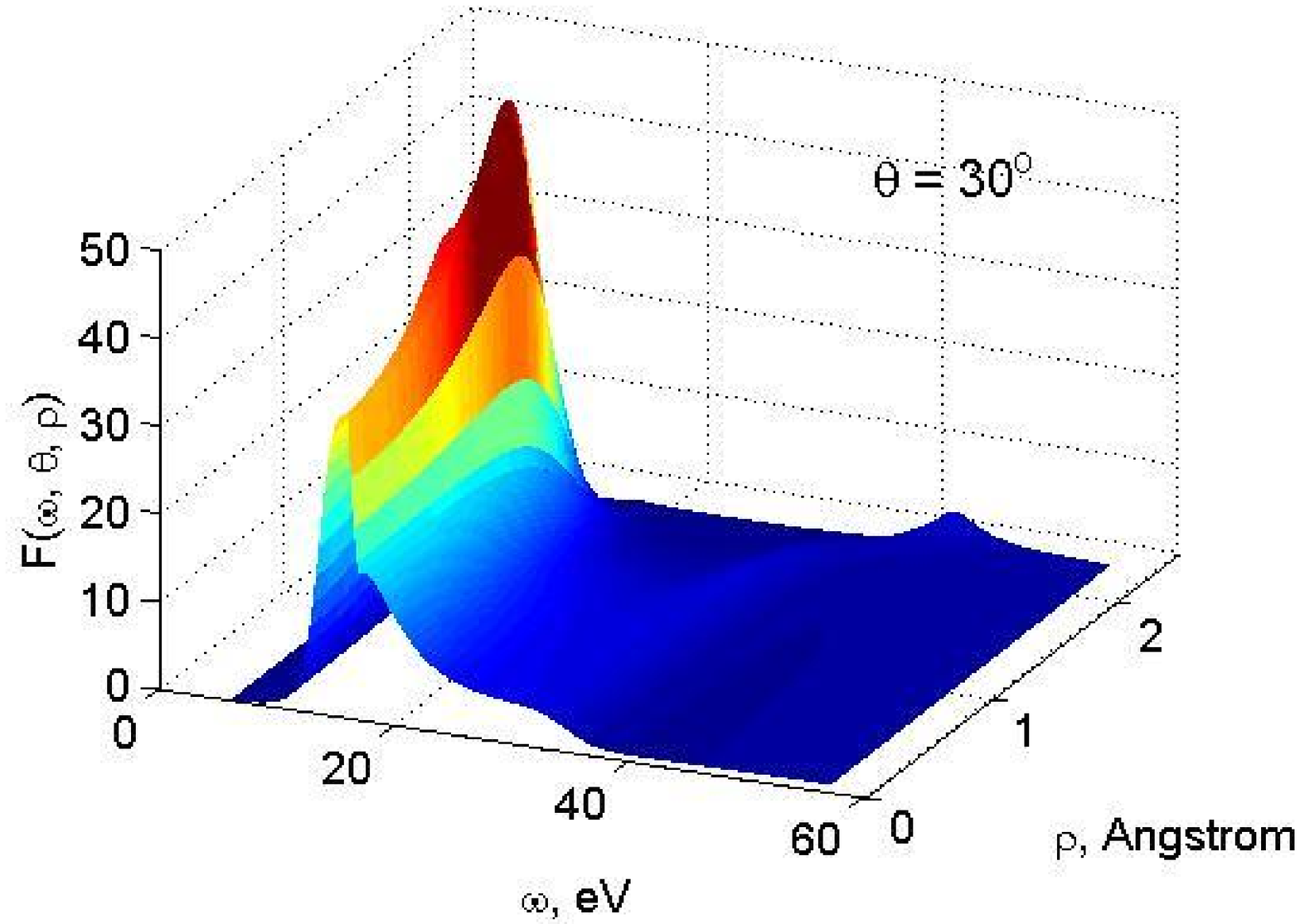}\\
\includegraphics[width=0.45\textwidth]{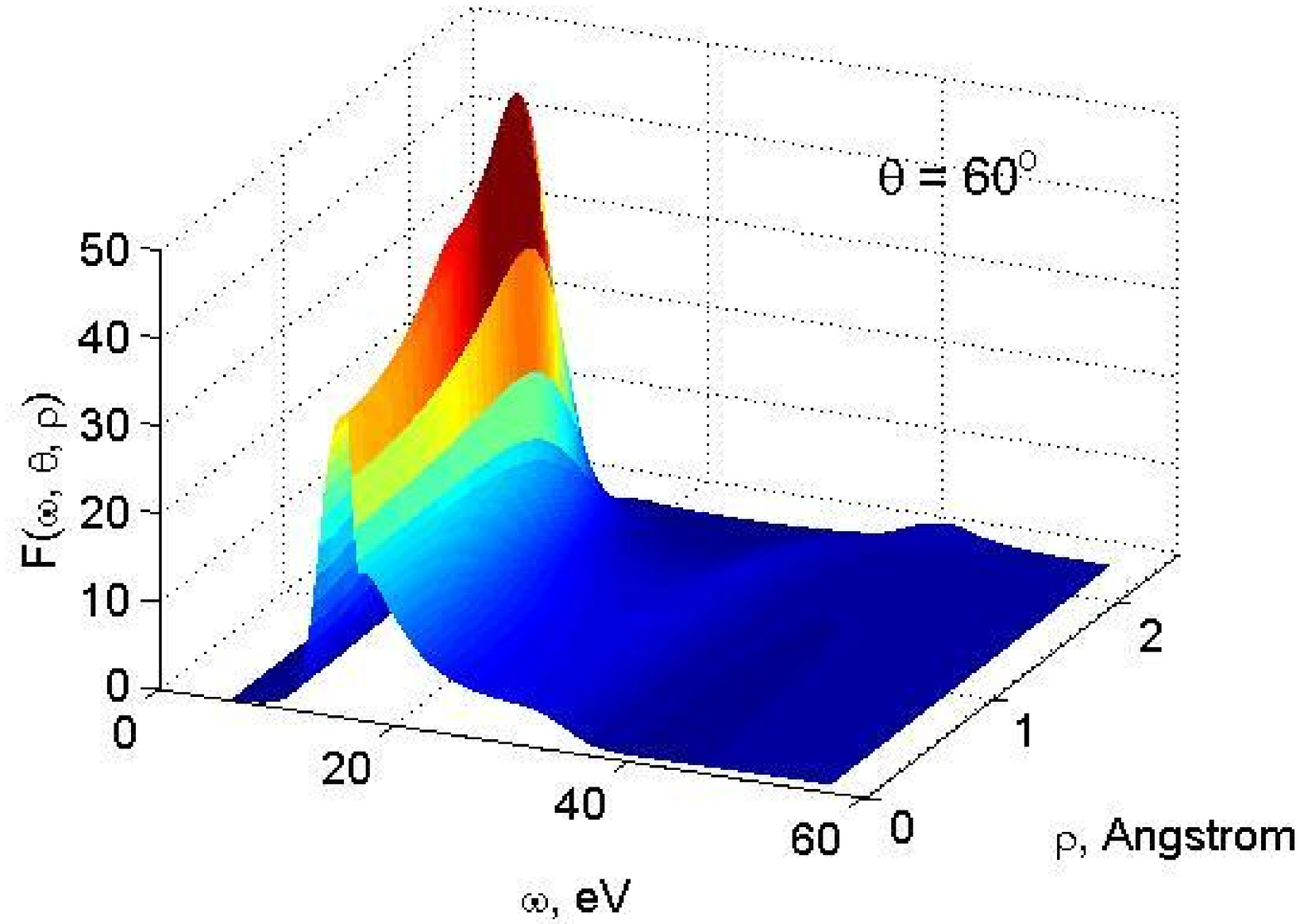}
\includegraphics[width=0.45\textwidth]{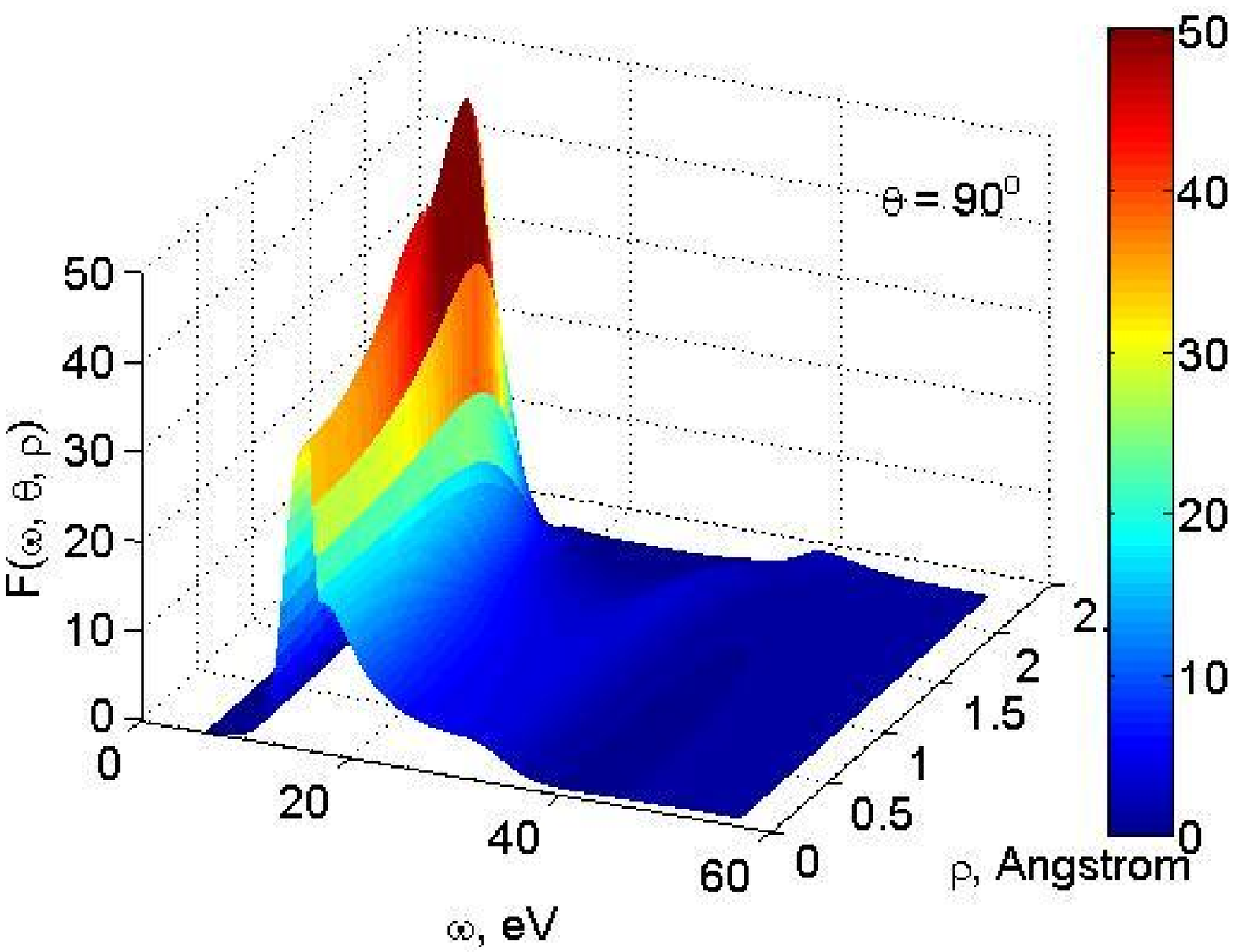}
\caption{The dynamical screening factor for Ar@C$_{60}$ as a function of photon energy $\omega$ and the radial distance of the atom from the fullerene's centre, $\rho$, calculated for different values of the angle $\theta$ between $\bfrho$ and $\bfE_0$.
}
\label{fig:c60_factor}
\end{figure}

\begin{figure}[t]
\centering
\includegraphics[width=0.45\textwidth]{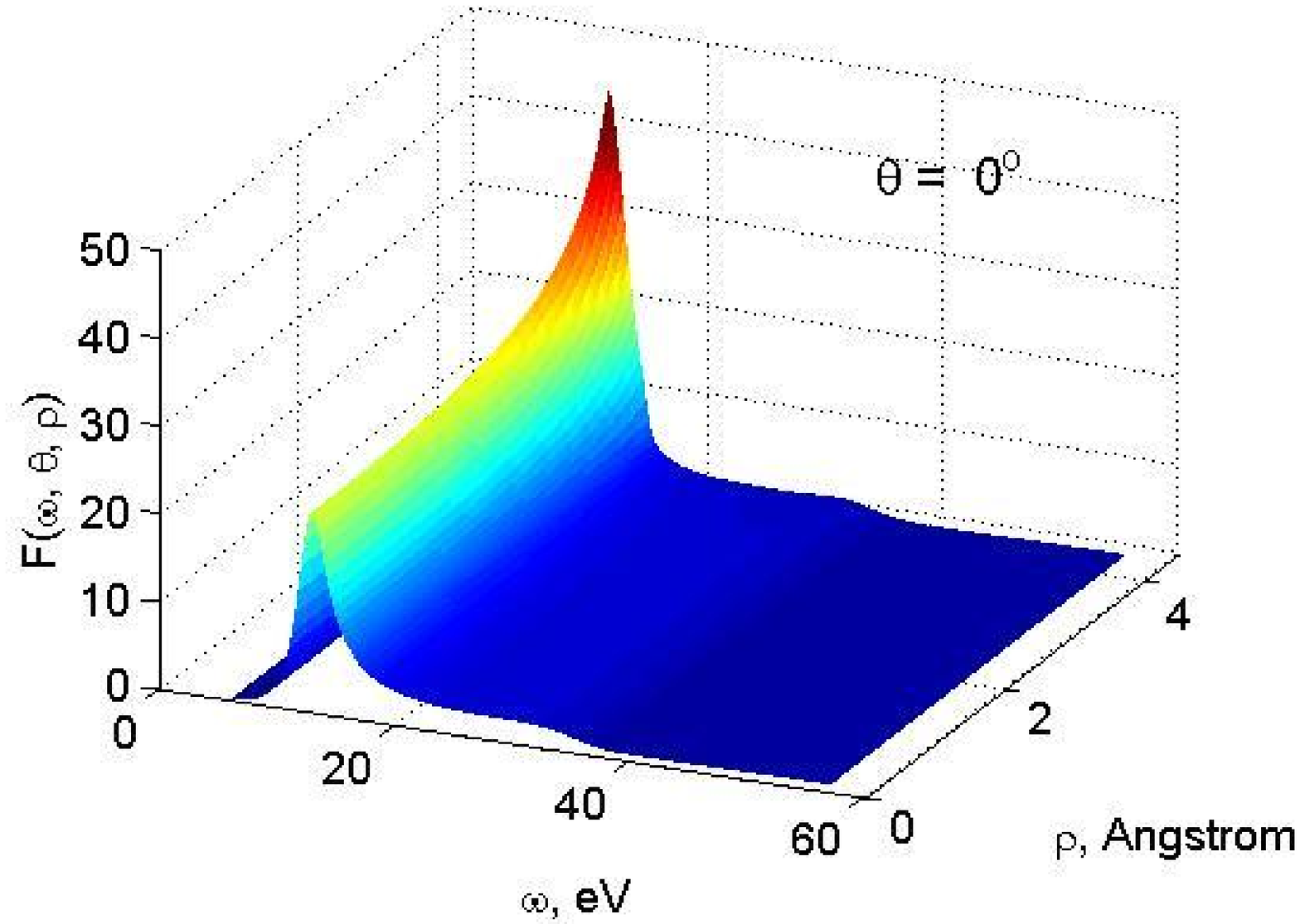}
\includegraphics[width=0.45\textwidth]{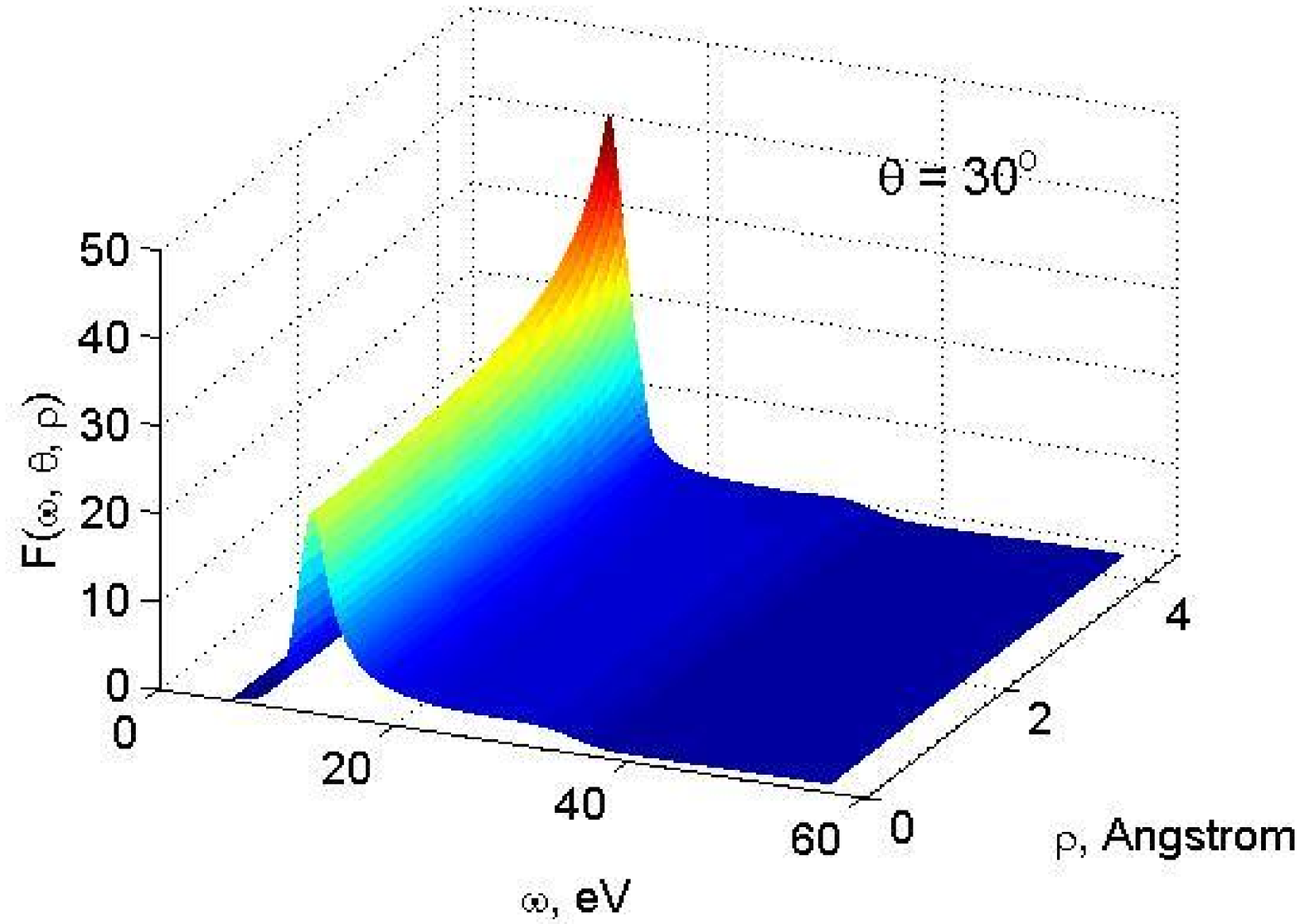}\\
\includegraphics[width=0.45\textwidth]{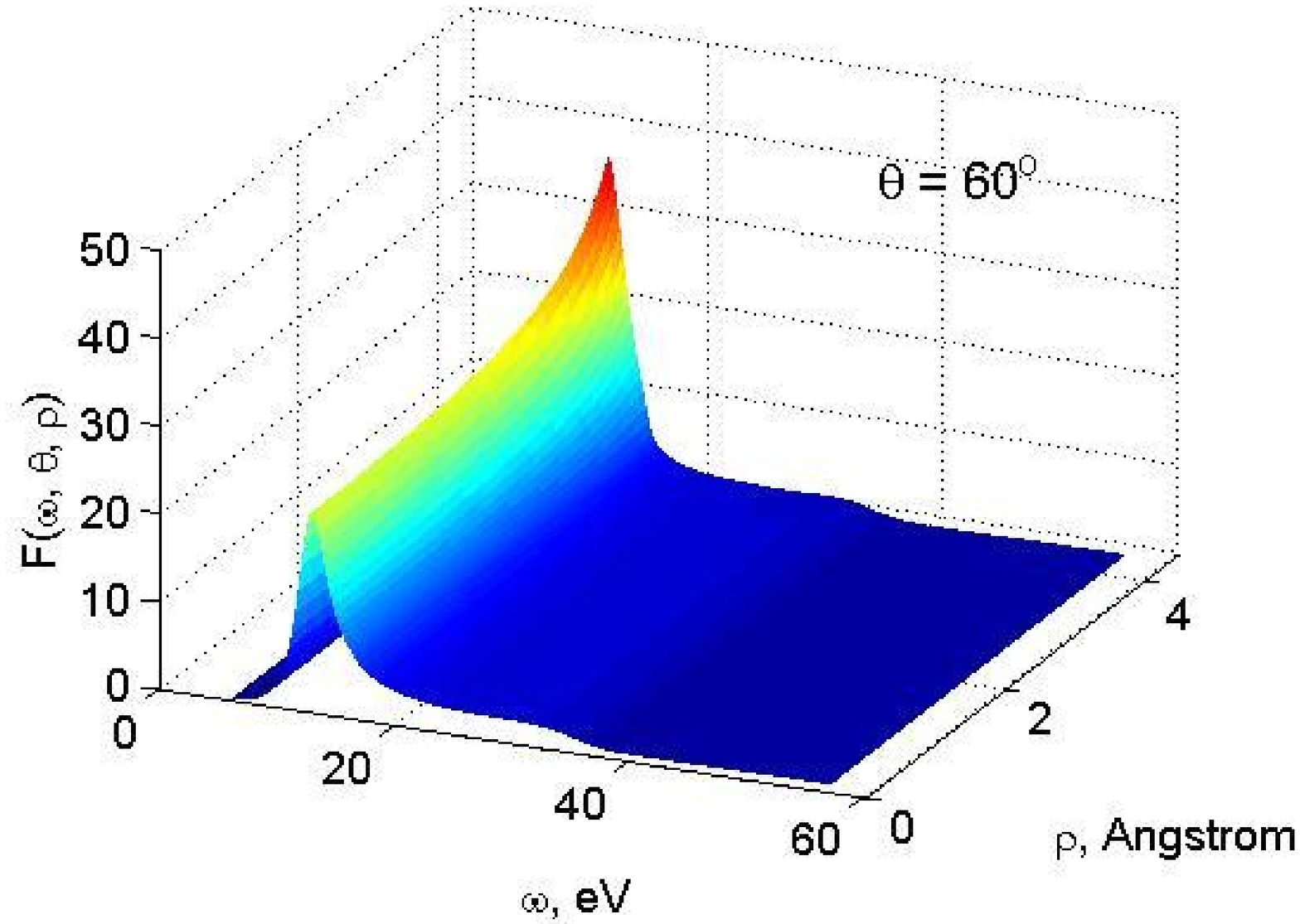}
\includegraphics[width=0.45\textwidth]{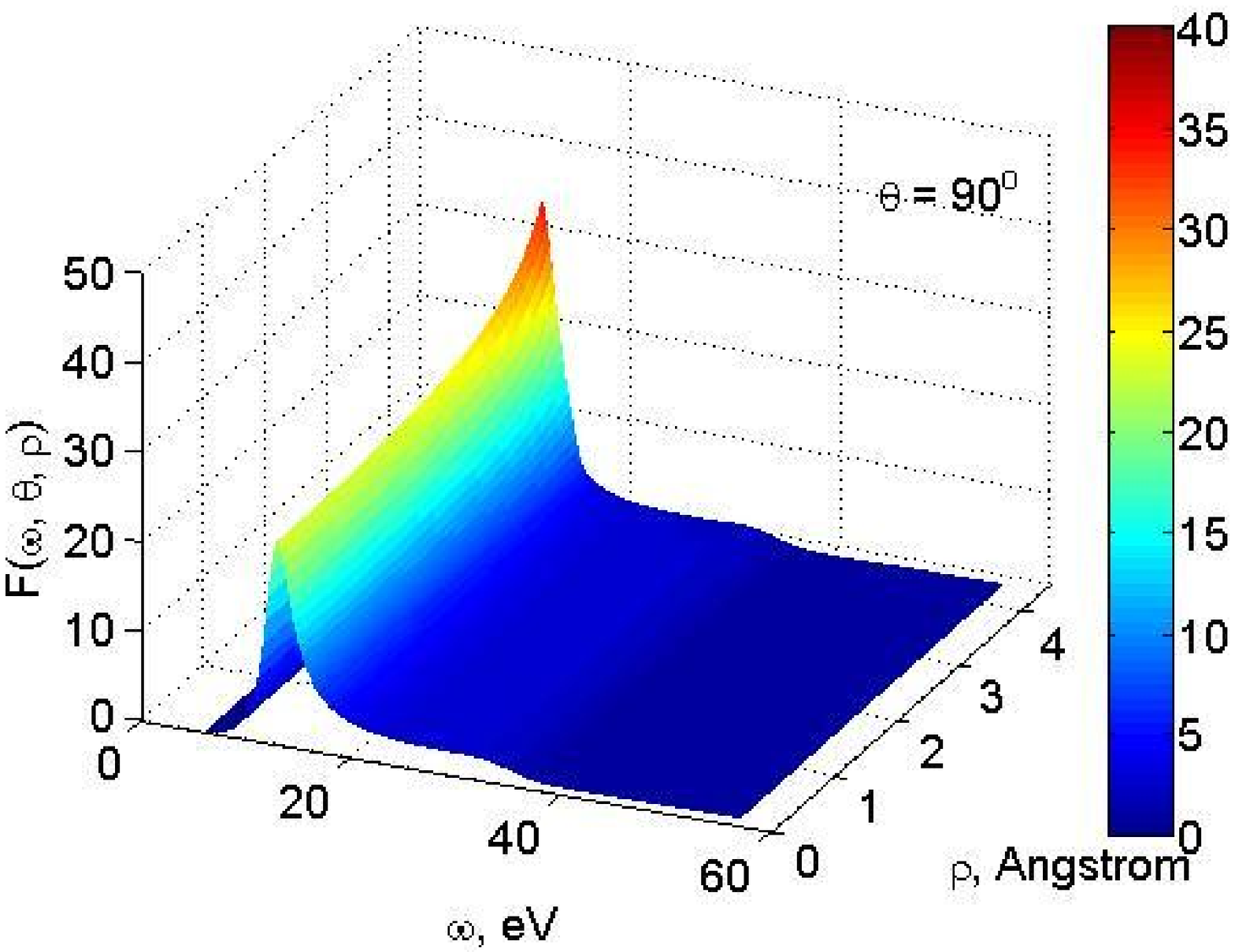}
\caption{Same as in figure \ref{fig:c60_factor} but for Ar@C$_{240}$.}
\label{fig:c240_factor}
\end{figure}

We illustrate the spatial dependence of the screening factor with two endohedral species Ar@C$_{60}$ and Ar@C$_{240}$.
As discussed in the previous section, there is a good quantitative agreement for the central case for ArC$_{60}$ between the unmodified model (\ref{eq:factor}) and the TDLDA calculation of \cite{MadjetChakraborty07}.
Therefore, there is no need to distinguish between the two types of delocalised electrons and the fullerene is modelled as a single spherical shell.
The factor for an arbitrarily positioned atom is calculated with (\ref{eq:factor}) for argon encapsulated inside C$_{60}$ and C$_{240}$ and is presented in figures \ref{fig:c60_factor} and \ref{fig:c240_factor}, respectively.

In each figure there are 4 panels showing the screening factor (\ref{eq:factor}) as a function of photon energy, $\omega$, and radial distance from the centre $\rho$ (see figure \ref{fig:noncentral_diagram}).
Each graph corresponds to the indicated value of the angle $\theta$ between $\bfrho$ and $\bfE_0$.

The mean radius $R$ of the C$_{240}$ cage is 7.1 \AA\, \cite{LuYang94}, just over twice that of the C$_{60}$ cage (3.5 \AA\, \cite{RuedelHentges02}).
The thickness of both fullerenes were set to 1.5 \AA.
The plasmon energies are calculated from (\ref{eq:plasmon_freq}).
In table \ref{tab:parameters}, the dipole surface plasmon energies ($l=1$) are given together with other relevant quantities that are used in this investigation.
The widths of all the plasmons are parameterized as $\Gamma_{jl} = 0.25 \omega_{jl}, j = 1,2$.
\begin{table}[t]
\caption{A table of the parameters of C$_{60}$ and C$_{240}$ used for evaluating (\ref{eq:factor}).
The various radii of the fullerenes are given in Angstroms, with the thickness of each fullerene set to 1.5 \AA.
The plasmon energies (in eV) are for the dipole surface plasmon mode only.}
\label{tab:parameters}
\begin{indented}
\item[]
\begin{tabular}{@{}lllllll}
\br
& $R$ & $R_1$  & $R_2$ & $\xi= R_1/R_2$ & $\omega_1$  & $\omega_2$  \\
\mr
C$_{60}$ & 3.5 \cite{RuedelHentges02} & 2.75 & 4.25 & 0.65 & 16.9 & 33.5 \\
C$_{240}$ & 7.1\cite{LuYang94} & 6.35 & 7.85 & 0.81 & 12.8 & 35.0 \\
\br
\end{tabular}
\end{indented}

\end{table}

The size of the fullerene cage has a significant impact on the dynamical screening factor.
Firstly, as discussed in \cite{LoKorol07}, the size of the cage determines the positions of the two plasmon modes.
With a larger shell, the two plasmon modes are positioned further apart, which in turn controls the corresponding widths.
This leads to differences in the energy dependence of the screening factors.
Secondly, the radius of the fullerene shell, together with the dynamic dipole polarizability of the atom, determines the strength of the interaction between the atom and fullerene and how much the screening factor is altered.
For Ar@C$_{60}$, this effect is strong because of the proximity of the atom to the fullerene cage.
For Ar@C$_{240}$, the much larger radius renders this effect negligible.
It is this second case that causes the most prominent differences between the screening factors of the two systems.

From the sets of graphs it can be seen that there is a rather weak dependence on $\theta$.
The radial distance from the centre, $\rho$, has a much larger influence on the dynamical screening factor, particularly at large distances.
As the atom is moved away from the centre of the fullerene, the screening factor (in the vicinity of the dipole symmetric plasmon mode), increases.
This increase becomes more dramatic as the atom approaches the shell, as can be seen in figure \ref{fig:c60_factor}.
In figure \ref{fig:c240_factor}, $\rho$ is only shown up to 4.3 \AA.
In this region, one can see the beginning of a steep rise in the magnitude of the screening factor.
Near the antisymmetric plasmon mode, the screening factor also increases, but only slightly.
In the energy range between these two modes, there are small decreases as the atom is moved away from the centre.
On the other hand, near the central region, the screening factor does not vary greatly.

\subsection{Spatial averaging for the screening factor}
As already mentioned, the optimal position of the endohedral atom is not necessarily at the centre of the fullerene.
Additionally, due to thermal motion the atom is not fixed at any one position inside the fullerene.
For meaningful observable results for the dynamical screening factor, it is necessary to perform an averaging over all possible positions of the endohedral atom. 

A spatial average of the dynamical screening factor $\langle\cF(\omega)\rangle$ can be carried out as follows:
\begin{equation}
\label{eq:average}
\langle \cF(\omega)\rangle
= 
%A 
\int_V F(\bfrho,\omega) \, W_{\mathrm{T}}(\bfrho) \, \mathrm{d}\bfrho.
\end{equation}
This integration is carried out over the volume $V$ where the atom may reside: a sphere of radius $R_1-R_a$, where $R_a$ is related to the size of the confined atom.
When the atom is sufficiently close to the fullerene cage, quantum effects such as hybridization become significant and the suitability of this semi-classical model is reduced.

The Boltzmann distribution $W_{\mathrm{T}}(\bfrho)$, governing the probability of finding the atom at $\bfrho$ and temperature $T$, is given by:
\begin{equation}
\label{eq:weight}
W_{\rm{T}}(\bfrho)
=
A \, 
\exp \left( - \frac{U_{\mathrm{tot}}(\bfrho)-U_{\mathrm{tot}}(0)}{kT} \right),
\end{equation}
where $k$ is the Boltzmann factor and $A$ is the normalisation constant obtained from $\int W_{\rm{T}}(\bfrho) \rd \bfrho = 1$.

The total interaction between the fullerene and the confined atom can be presented as a sum of two terms:
\begin{equation}
U_{\mathrm{tot}}(\bfrho) = U(\bfrho) + U_{\mathrm{vdW}}(\bfrho).
\label{eq:utot}
\end{equation}
The term, $U(\bfrho)$, is the potential energy of the interaction of the atomic dipole moment $\bfd(\bfrho)$ with the total electric field $\bfE_{\mathrm{tot}}(\bfrho)$:
\begin{equation}
\label{eq:interaction}
U(\bfrho) = -\frac{1}{2} \rm{Re}
\left(
\bfE_{\mathrm{tot}}^*(\bfrho) \bfd(\bfrho)
\right)
\end{equation}
The second term in (\ref{eq:utot}), $U_{\mathrm{vdW}}$, stands for the van der Waals interaction between the fullerene and the endohedral atom.
This is approximated by the Lennard-Jones potential, and is given by:
\begin{equation}
U_{\mathrm{vdW}}(\bfrho)
=
\sum_{i} \left( \frac{C_{12}}{(\bfr_i-\bfrho)^{12}} -
\frac{C_6}{(\bfr_i-\bfrho)^6} \right),
\label{eq:vdW_atoms}
\end{equation}
where the sum is carried out over all atoms of the fullerene and $r_i$ is position of the $i$-th atom.
The parameters for the interaction between an atom of the fullerene and the argon atom are taken from \cite{OndrechenBerkovitch-Yellin81}:
the depth of the potential well is 0.005 eV and the equilibrium distance between the two atoms is 3.84 \AA.
This translates to $C_6 = 53.8$ a.u. and $C_{12} = 3.93\times 10^6$ a.u. respectively.

\begin{figure}
\centering
\includegraphics[width=0.45\textwidth]{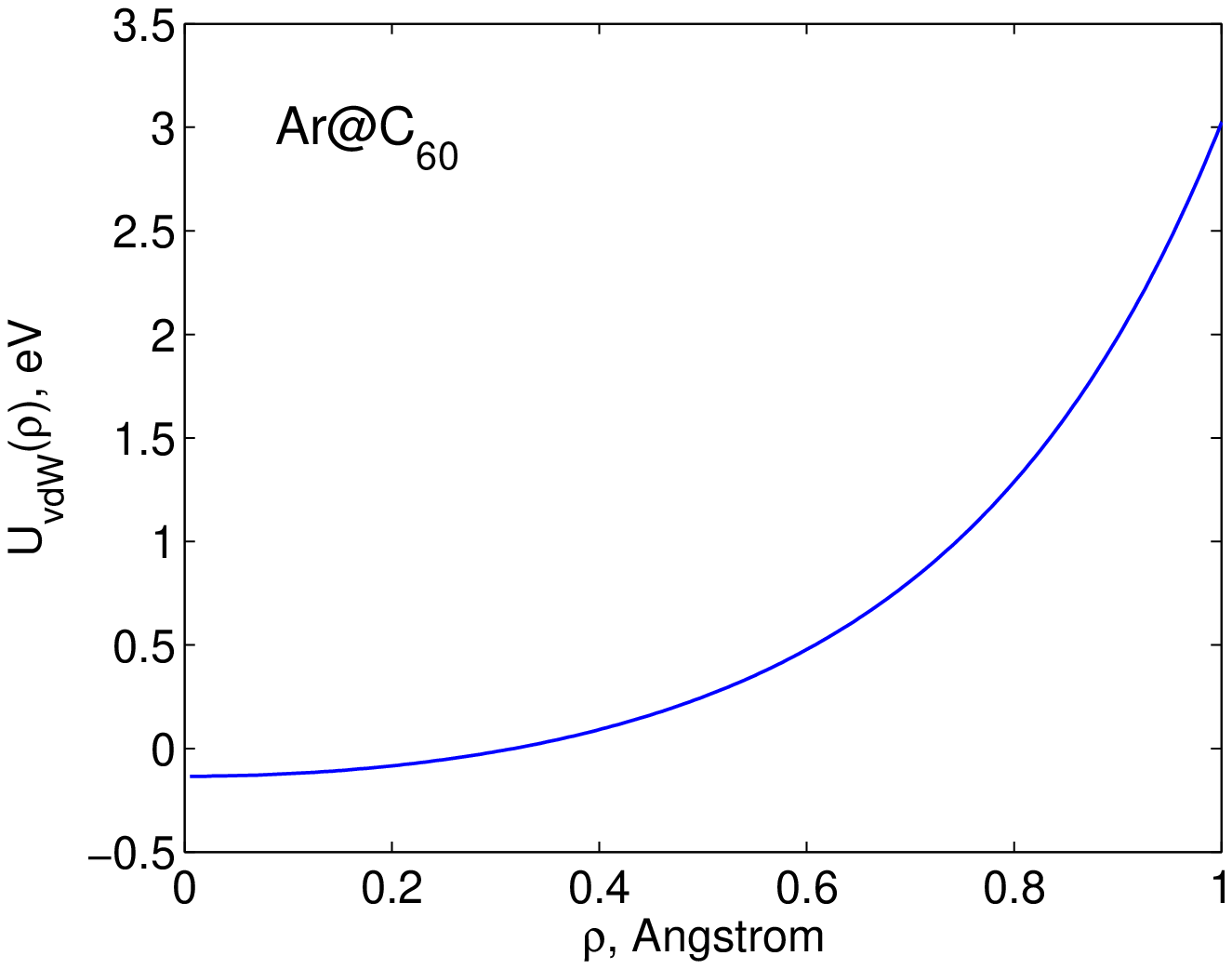}
\includegraphics[width=0.45\textwidth]{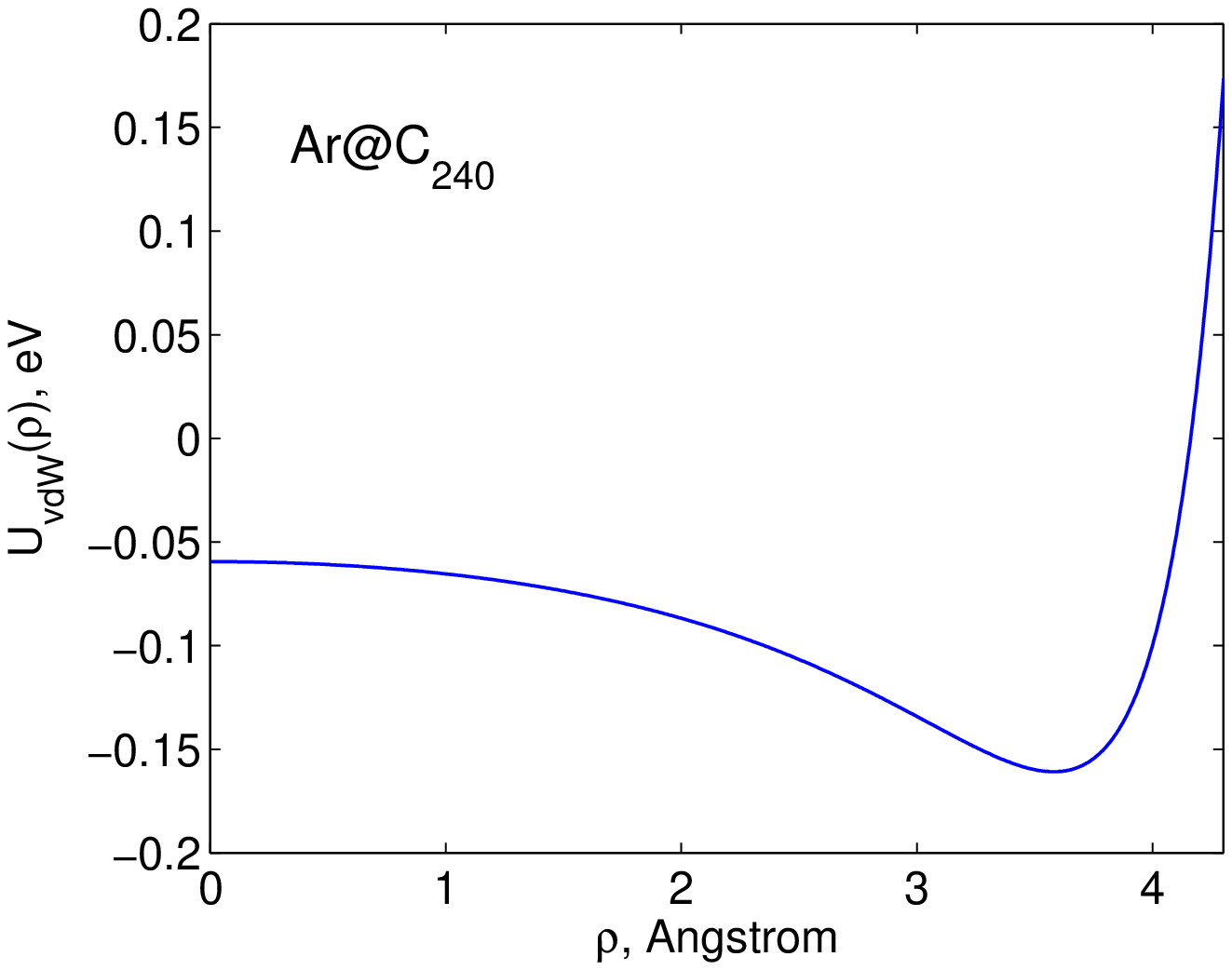}
\caption{The $U_{\mathrm{vdW}}$ profile for Ar@C$_{60}$ (left) and Ar@C$_{240}$ (right) are shown.}
\label{fig:vdW}
\end{figure}

For the evaluation of the van der Waals interaction energy, the fullerene is approximated as an infinitely thin sphere of radius $R$ and the sum of equation (\ref{eq:vdW_atoms}) becomes:
\begin{equation}
U_{\mathrm{vdW}}(\rho)
=
\int_S \left(
\frac{\sigma_{12}}{a^{12}} - \frac{\sigma_6}{a^6}
\right) \mathrm{d}S.
\label{eq:vdW}
\end{equation}
The integral is carried out over the surface of the sphere, eliminating the angular dependence of van der Waals interaction energy.
The distance between the surface element $\mathrm{d}S$ and the endohedral atom is denoted by $a$.
The quantities $\sigma_{6}$ and $\sigma_{12}$ for the fullerene of $n$ carbon atoms are defined as 
\begin{equation}
\sigma_{6}
=
\frac{n C_{6}}{4\pi R^2},
\qquad
\sigma_{12}
=
\frac{n C_{12}}{4\pi R^2},
\end{equation}
where $n/4\pi R^2$ is the surface density of carbon atoms.
Evaluating the integrals, one finds that the van der Waals interaction energy between the fullerene and the endohedral atom is
\begin{equation}
U_{\mathrm{vdW}}(\rho)
=
U_{12}(\rho) - U_6(\rho),
\end{equation}
where
\begin{eqnarray}
\cases{
U_{6}(\rho)
=
\frac{2\pi R \sigma_{6}}{4 \rho}
\left(
\frac{1}{(R-\rho)^4} - \frac{1}{(R+\rho)^4}
\right)
\\
U_{12}(\rho)
=
\frac{2\pi R \sigma_{12}}{10 \rho}
\left(
\frac{1}{(R-\rho)^{10}} - \frac{1}{(R+\rho)^{10}}
\right).}
\end{eqnarray}
The van der Waals potential profile for the two endohedral systems are shown in figure \ref{fig:vdW}.
\\

In both cases the dominant term in the total potential energy is the van der Waals term $U_{\mathrm{vdW}}(\bfrho)$.
In the case of Ar@C$_{60}$, the radius of the fullerene (3.5 \AA) is less than the equilibrium distance of the C$-$Ar van der Waals interaction (3.84 \AA).
With the exception of a small central region ($\rho<0.6$ \AA) the repulsive term of the Lennard-Jones potential dominates, as shown in the left panel of figure \ref{fig:vdW}.
The atom is, therefore, strongly confined to the centre of the fullerene.
For Ar@C$_{240}$, the size of the cage (7.1 \AA) is sufficiently large that there is a potential minimum, which lies in a spherical shell of radius of 3.6 \AA.
The depth of this potential well is 0.1 eV (see right panel of figure \ref{fig:vdW}).

\begin{figure}
\centering
\includegraphics[width=0.65\textwidth]{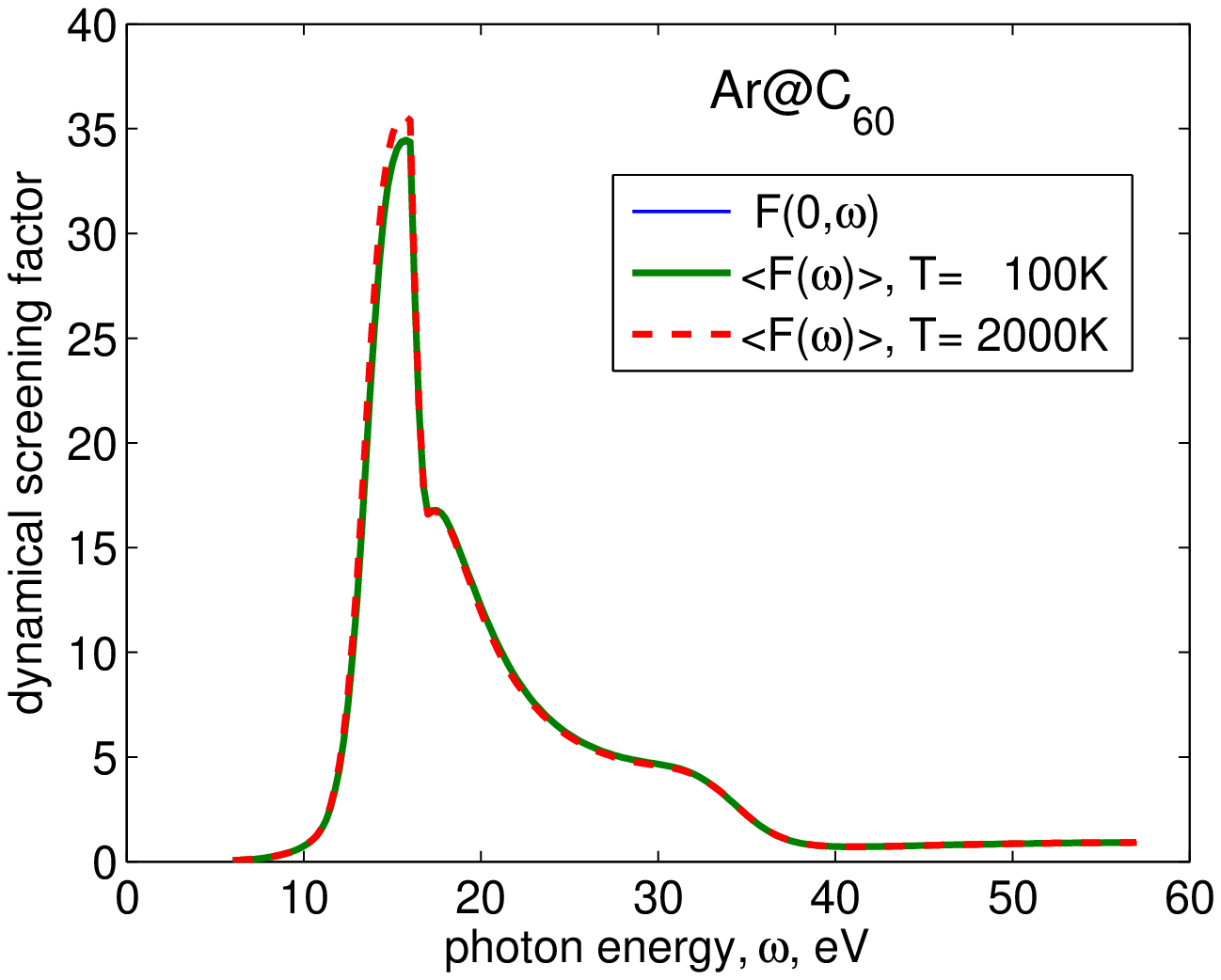}
\includegraphics[width=0.65\textwidth]{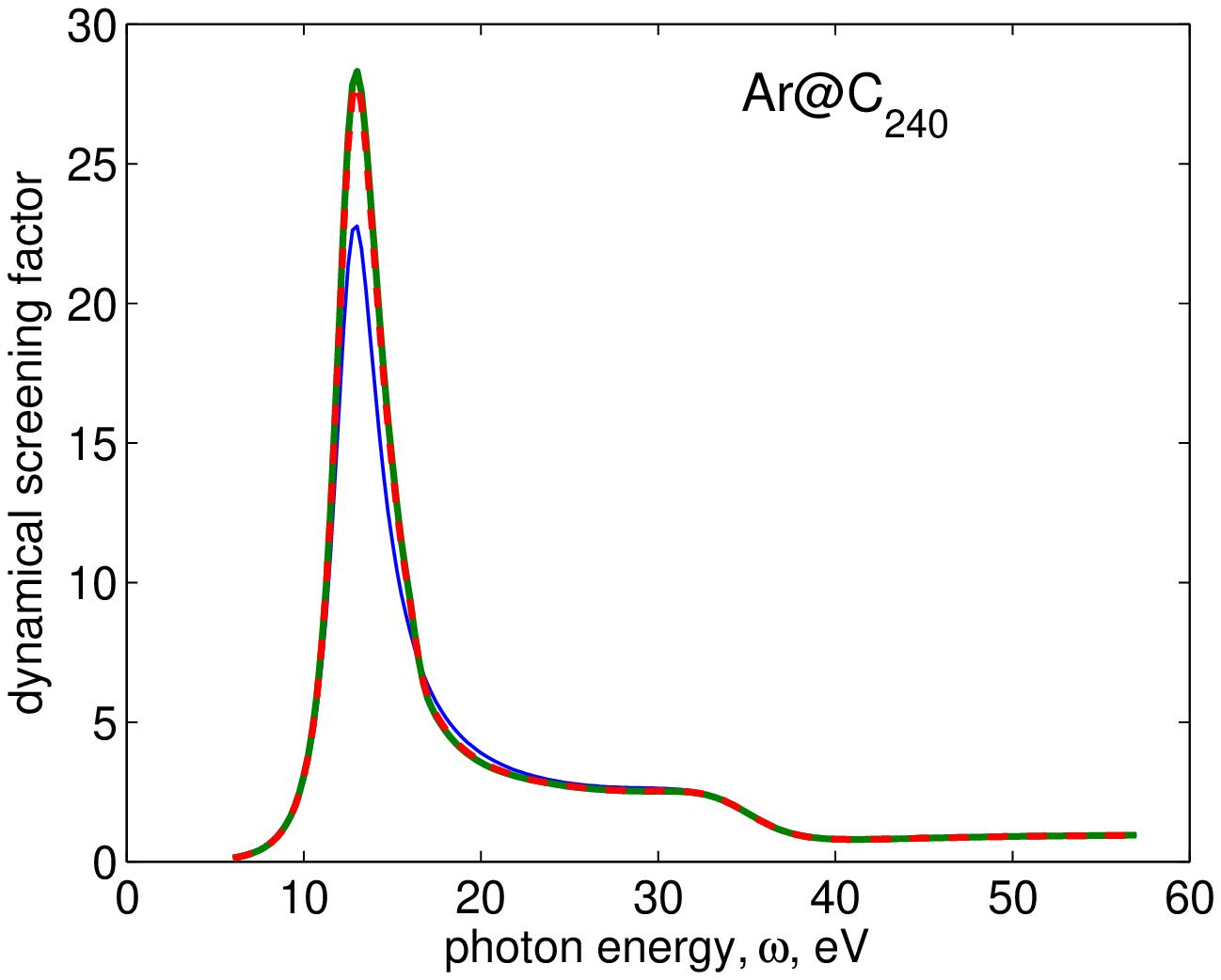}
\caption{The spatially averaged dynamical screening factor for Ar@C$_{60}$ (top) and for Ar@C$_{240}$ (bottom) for $T=100K$ (thick solid line) and $T=1500K$ (think dashed line).
The common legend is shown in the top panel.
The screening factor for the centrally positioned atom is shown for comparison (thin solid line).}
\label{fig:avg}
\end{figure}

The results of the spatial averaging are presented in figure {\ref{fig:avg}}.
In all cases, the external field strength was set to 0.001 a.u. (where 1 a.u. of electric field strength is $5 \times 10^9$ V/cm).
For Ar@C$_{60}$, the argon atom is essentially fixed at the centre.
The results of the spatial averaging reflects this.
Large increases in temperature only allows a minimal increase in the mobility of the endohedral atom, and has almost no impact on the spatially averaged dynamical screening factor.
For Ar@C$_{240}$, the argon atom should have enough thermal energy at $T \approx 1000 K$ to escape the potential well and reach the central region of the fullerene.
However, no significant changes are observed in the spatially averaged dynamical screening factor as the temperature is changed from $100 K$ up to $1500 K$.

\section{Summary}
In this semi-classical approach, a dielectric formalism was taken to study the dynamical screening of the endohedral atom.
The mutual interaction between the polarized atom and the polarized fullerene is now accounted for.
Previous work \cite{LoKorol07} has been generalised to consider an atom confined at an arbitrary position inside the fullerene.

A more indepth study of the central case was presented, showing the effect of accounting for the polarizability of the endohedral atom.
Case studies were performed on Ar@C$_{60}$ and Mg@C$_{60}$.
We show that the dynamical screening factor can depend strongly on the type of endohedral atom.

A comparison of the dynamical screening factor for Ar@C$_{60}$ and Mg@C$_{60}$ (the atom was positioned at the central position) was performed with the TDLDA calculations \cite{MadjetChakraborty07,ChakrabortyMadjet08}.
A good agreement was found for argon, however, only a qualitative agreement was observed for magnesium.

In an attempt to explain the discrepancies between our model and the TDLDA calculation, the dynamical screening factor was modified to include the two types of plasmons of the fullerene, the $\pi$ and the $\sigma$ plasmons.
This was achieved by modelling the fullerene as a system of two co-centric spherical shells, each being associated with one of the types of plasmons.
The modified model shows an improved correspondence with the TDLDA results.

Through the endohedral species Ar@C$_{60}$ and Ar@C$_{240}$, we show the spatial dependence of the dynamical screening factor.
There is only a weak dependence on the angle between $\bfrho$ and $\bfE_0$.
The distance between the endohedral atom and the fullerene cage has a significant impact on the screening factor - as the atom approaches the cage, its magnitude increases rapidly.

The size of the cage also impacts the energy dependence of the screening factor.
The response of the fullerene to the external field - the resonance energies of the symmetric and antisymmetric dipole plasmon modes - is controlled in part by its radius.
This was discussed in \cite{LoKorol07}.
In the present paper, the radius of the shell has an additional effect - it determines how strongly the screening factor is modified by the interaction between the atom and the fullerene.
For larger fullerenes, this interaction is diluted by the increased distance between the atom and the cage.

As the atom is not fixed to any position within the fullerene, a Boltzmann distribution was chosen to govern the probability of the confined atom being located at some arbitrary position within the fullerene.
This distribution is based on the interaction potential energy (electrostatic and van der Waals) between endohedral and the cage.
For the two cases studied, the van der Waals interaction is the dominant term. 
The argon atom is of similar size to the C$_{60}$ cage and is therefore located at the centre.
In the case of C$_{240}$, the cage is sufficiently large for the atom to move freely within it.
The van der Waals interaction between argon and the cage confines the atom to a spherical shell of radius of 3.6 \AA. 
In both cases, the spatially averaged dynamical screening factor is only weakly dependent on the temperature.

Despite the simplicity of this model, it provides an instructive description of the phenomena discussed here.
This model may be further applied to study the dynamic responses of more complicated situations such as multi-shelled fullerenes.

%\section*{Acknowledgements}
\ack
The authors are grateful for the helpful discussions with Prof. W. Greiner and Dr. A. Lyalin.

This work was supported by the  European Commission within the Network of Excellence project EXCELL (project number 515703) and by PECU under the grant 004916(NEST).

\section*{References}
%\bibliographystyle{unsrt}
%\bibliography{../papers/references}

\begin{thebibliography}{10}

\bibitem{LoKorol07}
Lo S, Korol A~V and  Solov'yov A~V
2007 {\em J. Phys. B: At. Mol. Opt. Phys.} {\bf 40} 3973--81

\bibitem{MadjetChakraborty07}
Madjet M~E, Chakraborty H~S and Manson S~T
2007 {\em Phys. Rev. Lett.} {\bf 99} 243003

\bibitem{ChakrabortyMadjet08}
Chakraborty H~S, Madjet M~E, J-M Rost and Manson S~T
2008 {\em Phys. Rev. A} {\bf 78} 013201

\bibitem{Sliwa96}
\'Sliwa W
1996 {\em Transition Metal Chemistry} {\bf 21} 583--92

\bibitem{Shinohara00}
Shinohara H
2000 {\em Reports on Progress in Physics} {\bf 63} 843--92

\bibitem{Harniet02}
Harneit W
2002 {\em Phys. Rev. A} {\bf 65} 032322

\bibitem{BenjaminArdavan06}
Simon~C Benjamin {\emph et al}
2006 {\em Journal of Physics: Condensed Matter} {\bf 18} S867--83

\bibitem{Tomanek05}
Tomanek D
2005 {\em J. Phys.: Condens. Matter} {\bf 17} R413--59

\bibitem{FatourosCorwin06}
Panos~P Fatouros {\em et al}
2006 {\em Radiology} {\bf 240} 756--64

\bibitem{ConneradeSolovyov05}
Connerade J-P and Solov'yov A~V
2005 {\em J. Phys. B: At. Mol. Opt. Phys.} {\bf 38} 807--13

\bibitem{LoKorolISACC08}
Lo S, Korol A~V and  Solov'yov A~V
2008 {\em Latest Advances in Atomic Cluster Collisions: Structure and Dynamics from the Nuclear to the Biological Scale}
Connerade J-P and Solov'yov A~V (ed.)
(London: Imperial College Press), p~162--176

\bibitem{Baltenkov99}
Baltenkov A~S
1999 {\em J. Phys. B: At. Mol. Opt. Phys.} {\bf 32}
2745--51

\bibitem{ConneradeDolmatov00}
Connerade J~P, Dolmatov V~K and  Manson S~T
2000 {\em J. Phys. B: At. Mol. Opt. Phys.} {\bf 33}
2279--85

\bibitem{AmusiaChernysheva02}
Amusia M~Ya, Chernysheva L~V, Manson S~T, Msezane A~M and Radojevi\'{c} V
2002 {\em Phys. Rev. Lett.} {\bf 88} 093002

\bibitem{KivimakiHergenhahn00}
Kivim\"aki A {\it et al}
%Hergenhahn U, Kempgens B, Hentges R, Piancastelli M~N, Maier K, R\"udel A, Tulkki J~J, and Bradshaw A~M
2000 {\em Phys. Rev. A} {\bf 63} 012716

\bibitem{AmusiaBaltenkov07}
Amusia M~Ya, Baltenkov A~S and Chernysheva L~V
2007 {\em Phys. Rev. A} {\bf 75} 043201

\bibitem{AmusiaBaltenkov08}
Baltenkov A~S, Amusia M~Ya, and Chernysheva L~V
2008 {\em Central Eur. J. Phys.} {\bf 6} 14--25

\bibitem{DolmatovManson08}
Dolmatov V~K and Manson S~T
2008 {\em J. Phys. B: At. Mol. Opt. Phys.} {\bf 41} 165001

\bibitem{DolmatovBrewer08}
Dolmatov V~K, Brewer P and Manson S~T
2008 {\em Phys. Rev. A} {\bf 78} 013415

\bibitem{LambinLucas92}
Lambin Ph, Lucas A~A and Vigneron J-P
1992 {\em Phys. Rev. B} {\bf 46} 1794--803

\bibitem{OestlingApell93a}
\"{O}stling D, Apell P and Ros\'{e}n A
1993 {\em Europhysics Letters} {\bf 21} 539--44

\bibitem{OestlingApell96}
\"{O}stling D, Apell P, Mukhopadhyay G and Ros\'{e}n A
1996 {\em J. Phys. B: At. Mol. Opt. Phys.} {\bf 29} 5115--25

\bibitem{AndersenBonderup00}
Andersen J~U and Bonderup E
2000 {\em Eur. Phys. J. D} {\bf 11} 413--34

\bibitem{Sihvola06}
Sihvola A~H
2006 {\em Progress In Electromagnetics Research (PIER)} {\bf 62} 317--31

\bibitem{BohrenHuffman}
Bohren C~F and Huffmann D~R
1998 {\em Absorption and Scattering of Light by Small Particles} (New York: Wiley)

\bibitem{Landau8}
Landau L~D and Lifshitz E~M
1960 {\em Course of Theoretical Physics} vol~8 {\em Electrodynamics of Continuous Media} (Oxford: Pergamon)

\bibitem{Henke93}
Henke B~L, Gullikson E~M and Davis J~C
1993 {\em Atomic data and nuclear data tables} {\bf 54} 181--342

\bibitem{FullereneAtlas}
Fowler P~W and Manolopoulos D~E
1995 {\em An Atlas of Fullerenes} ({\em International Series of Monographs on Chemistry} vol 30) (Oxford: Oxford University Press)

\bibitem{Solovyov05}
Solov'yov A~V
2005 {\em Int. J. Modern Phys. B} {\bf 19} 4143--84

\bibitem{IvanovKashenock01}
Ivanov V~K, Kashenock G~Yu, Polozkov R~G and Solov'yov A~V
2001 {\em J. Phys. B: At. Mol. Opt. Phys.} {\bf 34} L669--77

\bibitem{ReinkoesterKorica04}
Reink\"{o}ster A {\it et al}
%Korica S, Pr\"{u}mper G, Viefhaus J, Godehusen K, Schwarzkopf O, Mast M and Becker U
2004 {\em J. Phys. B: At. Mol. Opt. Phys.} {\bf 37} 2135--44

\bibitem{ScullyEmmons05}
Scully S~W~J {\it et al}
%Emmons E~D, Gharaibeh M~F, Phaneuf R~A, Kilcoyne A~L~D, Schlachter A~S, Schippers S, Muller A, Chakraborty H~S, Madjet M~E and Rost J~M
2005 {\em Phys. Rev. Lett.} {\bf 94} 065503

\bibitem{KorolSolovyov07}
Korol A~V and Solov'yov A~V
2007 {\em Phys. Rev. Lett.} {\bf 98} 179601

\bibitem{LuYang94}
Lu J~P and Yang W
1994 {\em Phys. Rev. B} {\bf 49} 11421--4

\bibitem{RuedelHentges02}
R\"udel A, Hentges R, Becker U, Chakraborty H~S, Madjet M~E and Rost J~M
2002 {\em Phys. Rev. Lett.} {\bf 89} 125503

\bibitem{OndrechenBerkovitch-Yellin81}
Ondrechen M~J, Berkovitch-Yellin Z and Jortner J
1981 {\em J. Am. Chem. Soc.} {\bf 103} 6586--92

\end{thebibliography}

\end{document}